\documentclass[journal=jctcce,manuscript=article]{achemso}
\setkeys{acs}{usetitle=true}

\usepackage[version=3]{mhchem} % Formula subscripts using \ce{}

\usepackage{amsmath,amssymb}
\usepackage{graphicx,color,subfigure}
\usepackage{amsfonts}
\usepackage{comment}
\usepackage{placeins}
\usepackage{fixltx2e}

\usepackage{sectsty}
\usepackage{balance}
\usepackage{xspace}

\usepackage{epsfig,amssymb,amsmath,amsthm,amsfonts,amsbsy,mathrsfs}
\usepackage{graphicx,color}
\usepackage{amsmath}
\usepackage{amssymb}
\usepackage{epstopdf}
\usepackage{threeparttable}
\usepackage{graphicx} %eps figures can be used instead
\usepackage{lastpage}
\usepackage[format=plain,justification=raggedright,singlelinecheck=false,font=small,labelfont=bf,labelsep=space]{caption}

\allowdisplaybreaks

\usepackage[version=3]{mhchem} % Formula subscripts using \ce{}

\usepackage{indentfirst} % first line indent after a section
\usepackage{siunitx} % use angstrom
\usepackage{braket}
\usepackage{tikz}
\usetikzlibrary{quantikz} % draw quantum circuit
\usepackage{algpseudocode} % for pesudo-code
\usepackage{algorithm}

\usepackage{multirow} % for table

\newcommand{\bt}{\mathbf{\theta}}

\title[Gradient]
{Circuit-Efficient Qubit-Excitation-based Variational Quantum Eigensolver}

\author{Zhijie Sun}
\affiliation{Key Laboratory of Precision and Intelligent Chemistry, University of Science and Technology of China, Hefei, Anhui 230026, China} 

\author{Jie Liu}
\email{liujie86@ustc.edu.cn} 
\affiliation{Hefei National Laboratory, University of Science and Technology of China, Hefei 230088, China}

\author{Zhenyu Li}
\email{zyli@ustc.edu.cn} 
\affiliation{Key Laboratory of Precision and Intelligent Chemistry, University of Science and Technology of China, Hefei, Anhui 230026, China} 
\alsoaffiliation{Hefei National Laboratory, University of Science and Technology of China, Hefei 230088, China}

\author{Jinlong Yang}
\affiliation{Key Laboratory of Precision and Intelligent Chemistry, University of Science and Technology of China, Hefei, Anhui 230026, China}
\alsoaffiliation{Hefei National Laboratory, University of Science and Technology of China, Hefei 230088, China}

\begin{document}

\begin{abstract}
The wave function Ans\"atze are crucial in the context of the Variational Quantum Eigensolver (VQE). In the Noisy Intermediate-Scale Quantum era, the design of low-depth wave function Ans\"atze is of great importance for executing quantum simulations of electronic structure on noisy quantum devices. In this work, we present a circuit-efficient implementation of two-body Qubit-Excitation-Based (QEB) operator for building shallow-circuit wave function Ans\"atze within the framework of Adaptive Derivative-Assembled Pseudo-Trotter (ADAPT) VQE. This new algorithm is applied to study ground- and excited-sate problems for small molecules, demonstrating significant reduction of circuit depths compared to fermionic excitation-based and QEB ADAPT-VQE algorithms. This circuit-efficient algorithm shows great promise for quantum simulations of electronic structures, leading to improved performance on current quantum hardware.

%We use the simplified-QEB operator pool as the operator-adding criterion, and use single parameter delta-E as the initial value for all-parameters optimization. Numerically, we compare the performance of sQEB operator pool with QEB and FEB in the framework of Adaptive Derivative-Assembled Pseudo-Trotter (ADAPT) VQE.
\end{abstract}

\section{Introduction} \label{sec:Introduction}
%\begin{itemize}

    %\item background of quantum computing

        The state space of quantum many-body system grows exponentially with the size of the system, leading to an exponentially computational complexity in both memory and time when solving electronic structure problems on a classical computer.\cite{Dirac29} It is anticipated to use a quantum system to simulate another quantum system in an efficient manner, that is quantum computing - an idea  envisaged by Feynman in 1982 \cite{Fey82,CaoYudRom19,McASamEnd20}. Quantum phase estimation (QPE) is the first quantum algorithm designed to solve electronic structure problems on a quantum computer, with a potential exponential speedup\cite{kitaev95,abrams99,AspDutLov05}. While the QPE algorithm requires a large number of qubits and gates, which is impossible to implement on Noisy Intermediate-Scale Quantum (NISQ) devices\cite{Pre18,BhaCerKya22}. Alternatively, the hybrid quantum-classical algorithms, such as variational quantum eigensolver (VQE), have been proposed to reduce the circuit depth and meanwhile mitigate error in the presence of noise on NISQ devices~\cite{PerMcCSha14,McCJarRom16,WanHigBri19,HemMaiRom18,NamChePis20,SheZhaZha17,OMalley16,ColRamDah18,TilJulChe22,CerArrBab21,LiuLiYan21,LiuFanLi22,HuaCaiLi22,FedPenGov22,BauBelBra20}. The VQE algorithm uses quantum computers to prepare a parameterized quantum state and then measure the expectation value of the Hamiltonian. The circuit parameters that minimize the total energy are optimized on classical computers. Given the variational principle introduced in the VQE, the energy minimum is a upper bound to the exact energy of the target state.
    
    %\item adaptive variational quantum algorithms

        The wave function ansatz that is represented by a quantum circuit in the context of quantum computing is at the core of a VQE algorithm\cite{KanMezTem17,LeeHugHea18,RyaYenGen18,RyaLanCen20,XiaKai20,AnaSchAlp22,GarZhuBar20,FanLiuLi23,ZenFanLiu23}. A good wave function ansatz is required to have both powerful expressivity and shallow circuit depth, which are closely related to the accuracy and efficiency of quantum simulations, respectively. One of the most widely used ansatzes for electronic structure simulations is unitary coupled cluster (UCC)~\cite{PerMcCSha14,GriClaEco19,RomBabMcC18,BarGonSok18}, a unitary version of the traditional coupled cluster approach. Implementing the UCC ansatz on a quantum computer requires to perform a Trotter-Suzuki decomposition of the UCC cluster operator and then compile them into one- and two-qubit gates, which leads to Trotter error at a finite-order truncation. Adaptive Derivative-Assembled Pseudo-Trotter (ADAPT) VQE \cite{GriEcoBar19} is an alternative scheme to build a ``factorized'' form of the UCC ansatz and approach the exact wave function with an arbitrarily long product of exponentialized one- and two-body excitation operators. 
    
    %\item qubit variation quantum algorithm

        The ADAPT-VQE algorithm is an appealing scheme for simulating electronic structure on near-term quantum computers with a compact wave function ansatz. However, limited by fidelity of quantum gate operators, it is crucial to reduce the ansatz circuit depth in order to suppress error accumulation. For example, Tang et al. decomposed the fermionic excitation operator into Pauli strings and defined the operator pool with a set of Pauli strings with the length less than 4, known as a qubit version of ADAPT-VQE (qubit-ADAPT-VQE)~\cite{TanShkBar21}. In contrast to the original ADAPT-VQE algorithm, the qubit-ADAPT-VQE algorithm significantly reduces the number of controlled-NOT (CNOT) gates but increases the number of variational parameters.  Yordan et al. proposed a CNOT-efficient implementation for fermionic excitation-based (FEB) and qubit excitation-based (QEB) operators~\cite{YorArvBar20}, and combined the latter one with the ADAPT-VQE algorithm, as a compromising scheme to balance the number of CNOT gates and variational parameters~\cite{YorYorArm21}. The optimized quantum circuits for implementing single and double QEB operators require only 2 and 13 CNOT gates, respectively, while the convergence rate of QEB-ADAPT-VQE is comparable to that of FEB-ADAPT-VQE. Recently, Magoulas et al. generalized these CNOT-efficient quantum circuits to arbitrary excitation ranks for an efficient implementation of selected projective quantum eigensolver (SPQE)~\cite{MagEva23}. Furthermore, they proposed linear-scaling quantum circuits for approximately implementing high-order excitation operators without loss of accuracy in the SPQE simulations but at the expense of breaking the particle number and total spin projection symmetries~\cite{MagEva23b}. Although these CNOT-efficient algorithms provide promising schemes to build compact wave function ansatzes, finding an optimal wave function ansatz is still an open problem. 

    %\item introduce your method

        In this work, we proposed CNOT-efficient Qubit-Excitation-Based operators, named simplified QEB (sQEB), for quantum simulations of quantum chemistry. In contrast to the QEB operators, the sQEB operators are able to further reduce the number of CNOTs and meanwhile conserve the number of particles and the S$z$ component of the spin. We numerically compare the convergence rate and quantum resource requirement of sQEB-ADAPT-VQE with those of QEB-ADAPT-VQE and FEB-ADAPT-VQE by evaluating the dissociation curves of LiH, BeH$_2$ and H$_6$ chain, demonstrating that fewer CNOT gates are required to implement the sQEB-ADAPT-VQE for electronic structure simulations. We also successfully apply the sQEB-ADAPT-VQE to compute excited states of LiH and BeH$_2$.

        %The paper is organized as follows. in section II we introduce the basic theory and give the definition of the sQEB operators. In section III we give the numerical results for benchmarking the performance of sQEB. In Section VI we conclude our results.
%\end{itemize}

\section{Methodology} \label{sec:Methodology}

    \subsection{Adaptive variational quantum algorithms}

        The electronic Hamiltonian in second quantization is 
            \begin{equation}
                H = E_\mathrm{NN} + \sum_{pq} h_{pq} \hat{e}^{p}_{q}+\frac{1}{2}\sum_{pqrs} h_{pqrs} \hat{e}^{pq}_{rs},
            \end{equation}
        where 
        $\hat{e}^{p}_{q}=a^{\dag}_pa_q$ and $\hat{e}^{pq}_{rs}=a^{\dag}_pa^{\dag}_qa_ra_s$ are single and double excitation operators, and $h_{pq}$ and $h_{pqrs}$ is one- and two-electron integrals. $E_\mathrm{NN}$ is the nuclear repulsion energy. The ground-state problem is to solve the eigenvalue equation
        \begin{equation}
            H \ket{\psi} = E \ket{\psi}.
        \end{equation}

        To utilize quantum computers to solve the Schr\"odinger equation, one first needs to map it onto the qubit representation using Jordan-Wigner (JW) \cite{JorWig28}, Bravyi-Kitaev (BK) or parity transformation \cite{BraKit02, SeeRicLov12}. After the mapping, the Hamiltonian can be generally written as 
        \begin{equation}
            H = \sum_\mu h_\mu \hat{P}_\mu,
        \end{equation}
        where $\hat{P}_i$ is Pauli string of $\{I,X,Y,Z\}^{\otimes N}$. The ground-state wave function can be represented as 
        \begin{equation}
            \ket{\psi(\bt)}  = U(\bt) \ket{\psi_0},
        \end{equation}
        where $U(\bt)$ is a unitary transformation that is decomposed into product of a sequence of one- and two-qubit gates to implement on a quantum computer. $\ket{\psi_0}$ is the initial state, commonly chosen to be Hartree-Fock state $\ket{\psi_{hf}}$ for quantum simulations of electronic structure. 

        Adaptive variational quantum algorithms iteratively build a wave function ansatz in the form of
        \begin{equation}
            U(\bt) = \prod_{k=1}^{N_k} U(\theta_k), 
        \end{equation}
        where $U(\theta_k)$ is a unitary exponentialized operator in the form of $U(\theta_k)=\mathrm{exp}(\theta_k \tau_k)$ with $\tau_k$ being an anti-Hermitian operator. In each iteration, one needs to determine the operator $\tau_\mu$ with the largest absolute residual gradient 
        \begin{equation}\label{Gradient}
        \begin{split}
            G_\mu &= \left. \frac{\partial \bra{\psi_k} U^\dagger(\bt_\mu) H U(\bt_\mu) \ket{\psi_k}}{\partial \theta_\mu} \right |_{\theta_\mu=0}\\
            &= \bra{\psi_k} [H, \tau_\mu] \ket{\psi_k}
        \end{split}
        \end{equation}
        from the operator pool $\mathcal{O} = \{\tau_\mu\}$ and add it to the operator sequence $\{\tau_1,\cdots,\tau_k\}$. Then, one use classical computer to optimize the variational parameters $\bt$ to minimize the expectation value \begin{equation}
        E(\bt)=\bra{\psi_0}U^\dag(\bt)HU(\bt)\ket{\psi_0}.
        \end{equation}
        
        The ADAPT-VQE algorithm is described in Algorithm~\ref{alg:ADAPT}.
        
        \begin{algorithm}[H]
        \caption{ADAPT-VQE procedure}\label{alg:ADAPT}
        \begin{algorithmic}[1]
            \Procedure{ADAPT-VQE}{}
            \State Initialize the Hamiltonian $H$, the operator pool $\mathcal{O}$ and initial wave function $\ket{\psi_0}$.
            \While{$|\mathbf{G}|>\epsilon$}
                \State {\textbf for} $\tau_\mu$ in $\mathcal{O}$ 
                \State \quad Calculate the residual gradient $G_\mu$ using Eq.(\ref{Gradient}).
                \State {\textbf end}
                \State Determine $\tau_k$ with the largest absolute residual gradient.
                \State Update the wave function with $\ket{\psi_k} = U(\theta_k) \ket{\psi_{k-1}}$.
                \State Optimize variational parameters to minimize the energy $E(\bt)$.
            \EndWhile
            \State Output the wave function ansatz and the total energy.
            \EndProcedure
        \end{algorithmic}
        \end{algorithm}
    
    \subsection{Qubit excitation-based operator pool}

       In the ADAPT-VQE algorithm, the FEB operator pool $\{\hat{e}_p^q-\hat{e}_q^p, \hat{e}^{pq}_{rs}-\hat{e}^{rs}_{pq}\}$ or its spin-adapted version is often employed to iteratively build the wave function ansatz. After the JW transformation, the general anti-Hermitian fermionic excitation operators are mapped onto the qubit representation as 
       \begin{equation}
\begin{split}
    &\hat{e}^p_q - \hat{e}^q_p = (Q_p^\dagger Q_q - Q_q^\dagger Q_p) \prod^{q-1}_{k=p+1}{Z_k}\\
    &\hat{e}^{pq}_{rs} - \hat{e}^{rs}_{pq}= (Q_p^\dagger Q_q^\dagger Q_r Q_s - Q_s^\dagger Q_r^\dagger Q_q Q_p) \prod^{q-1}_{k=p+1}{Z_k}\prod^{s-1}_{l=r+1}{Z_l},
\end{split}
\end{equation}
Here $p < q < r < s$ and the qubit creation and annihilation operators are defined as
\begin{equation}
    \begin{split}
        Q^\dag = \frac{1}{2}(X-iY)\\
        Q = \frac{1}{2}(X+iY).
    \end{split}
\end{equation} 
Yordan et al. proposed CNOT-efficient circuits to represent FEB excitations, in which at least $2|p-q|+1$ and $2|q+s-p-r|+9$ CNOTs gates are required to encode one- and two-body fermionic excitation operators into quantum circuits, respectively~\cite{YorArvBar20}. 

        In order to further reduce the circuit depth, Yordan et al. suggested to remove the Pauli-Z strings from the fermionic excitation operators and thus defined the QEB operator pool
        \begin{equation}
            \{Q^{\dag}_p Q_q-Q^\dagger_q Q_p, Q^{\dag}_p Q^{\dag}_q Q_r Q_s-Q^{\dag}_s Q^{\dag}_r Q_q Q_p\}
        \end{equation}
        Here, the two-body QEB operator can be expanded as
        \begin{equation}\label{eq:qeb_op}
        \begin{split}
            \kappa^{pq}_{rs} = &Q^{\dag}_p Q^{\dag}_q Q_r Q_s-Q^{\dag}_s Q^{\dag}_r Q_q Q_p \\
            = - \frac i8 \{ &X_pY_qX_rX_s + Y_pY_qY_rX_s - X_pX_qX_rY_s - Y_pX_qY_rY_s \\
            + &Y_pX_qX_rX_s + Y_pY_qX_rY_s - X_pX_qY_rX_s - X_pY_qY_rY_s \},
        \end{split}
        \end{equation}
        The quantum circuits used to implement one- and two-body QEB operators only consist of 2 (see Fig.~\ref{fig:QEB_single_circuit}) and 13 CNOTs (see Fig.~\ref{fig:QEB_circuit}), respectively~\cite{YorYorArm21}. 
        
        \begin{figure}[H]
            \begin{center}
            \noindent\makebox[\textwidth]{
            \begin{quantikz}
                &\lstick{$p$} &\gate{R_z(\frac{\pi}{2})}  &\gate{R_x(\frac{\pi}{2})}  &\ctrl{1}  &\gate{R_x(\theta)} &\ctrl{1}  &\gate{R_x(-\frac{\pi}{2})}  &\gate{R_z(-\frac{\pi}{2})}         &\qw\\
                &\lstick{$q$} &\qw   &\gate{R_x(\frac{\pi}{2})}   &\targ{}  &\gate{R_z(\theta)}   &\targ{}  &\gate{R_x(-\frac{\pi}{2})}  &\qw    &\qw
            \end{quantikz}
            }\end{center}
            
            \caption{Quantum circuit for realizing the one-body QEB operator.}
            \label{fig:QEB_single_circuit}
        \end{figure}
        
        \begin{figure}[H]
            \begin{center}
            \noindent\makebox[\textwidth]{
            \begin{quantikz}
                &\lstick{$p$} &\ctrl{2} &\qw        &\targ{}   &\octrl{1} &\targ{}  &\qw       &\ctrl{2} &\qw\\
                &\lstick{$q$} &\qw       &\ctrl{2}  &\ctrl{-1}  &\gate{R_y(\theta)}   &\ctrl{-1}  &\ctrl{2}  &\qw       &\qw \\
                &\lstick{$r$} &\targ{}   &\qw       &\qw       &\ctrl{-1}                 &\qw        &\qw       &\targ{}   &\qw\\
                &\lstick{$s$} &\qw       &\targ{}   &\qw       &\ctrl{-1}                 &\qw        &\targ{}   &\qw       &\qw
            \end{quantikz}
            }\end{center}
            
            \caption{Quantum circuit for realizing the two-body QEB operator.}
            \label{fig:QEB_circuit}
            
        \end{figure}

Here, we employed a modified but equivalent circuit for realizing the two-body QEB excitation in contrast to the original one presented in Ref.~\citenum{YorArvBar20}. In fact, there exist many other quantum circuits for representing two-body QEB excitation operators, as detailed in the supplementary information. All of these circuits have similar structures that consist of computational basis rotation using multiple CNOTs and quantum state rotation using a multi-qubit controlled Ry (CRy) gate. As shown in Fig.~\ref{fig:QEB_circuit}, the 3-qubit controlled Ry gate carries out state rotation by mixing $\ket{1100}$ and $\ket{1101}$, while leaving the rest of computational basises unchanged. After the computational basis rotation, the two-body QEB circuit implements state rotation in the form of $\cos{\theta}\ket{0011} + \sin{\theta}\ket{1100}$. It is evident that the QEB excitation operators conserve both the number of particles, $N$, and the S$_z$ component of the spin if $\kappa^{pq}_{rs}$ satisfies the following condition
\begin{equation}
    \sigma_p+\sigma_q=\sigma_r+\sigma_s
\end{equation}
where $\sigma_i$ is the spin of the $i$-th spin-orbital, and $\sigma_i$ is 0(1) if the $i$-th spin-orbital is spin up(down). 

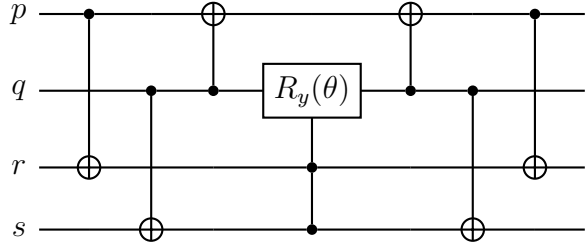
\begin{figure}[H]
            \begin{center}
            \noindent\makebox[\textwidth]{
            \begin{quantikz}
                &\lstick{$p$} &\ctrl{2} &\qw       &\targ{}   &\qw   &\targ{}   &\qw       &\ctrl{2} &\qw\\
                &\lstick{$q$} &\qw       &\ctrl{2} &\ctrl{-1} & \gate{R_y(\theta)} &\ctrl{-1} &\ctrl{2} &\qw       &\qw\\
                &\lstick{$r$} &\targ{}   &\qw       &\qw       &\ctrl{-1}            &\qw       &\qw       &\targ{}   &\qw\\
                &\lstick{$s$} &\qw       &\targ{}   &\qw       &\ctrl{-1}            &\qw       &\targ{}   &\qw       &\qw
            \end{quantikz}
            }\end{center}
            
            \caption{Quantum circuit for realizing the sQEB excitation operator.}
            \label{fig:sQEB_circuit}
        \end{figure}

In case of two-body qubit excitations, there are in principle four configurations, including  
\begin{equation}
 \ket{1100}, \ket{1001}, \ket{0110}, \ket{0011} 
\end{equation}
which satisfy conservation of $N$ and S$z$. This implies that one needs at least two controlled qubits in the multi-qubit CRy gate to construct CNOT-efficient circuits, which realize state rotation among these four configurations. One example is to keep state rotation between $\ket{1100}$ and $\ket{1101}$ unchanged and introduce new state rotation between $\ket{1110}$ and $\ket{1111}$, that is one removes the controlled qubit of $p$ as shown in Fig.~\ref{fig:sQEB_circuit}. 

The simplified QEB (sQEB) quantum circuit in Fig~\ref{fig:sQEB_circuit} can be represented as
        \begin{equation}        
            \tau^{pq}_{rs}
            = - \frac i4(X_pY_qX_rX_s+Y_pY_qY_rX_s-X_pX_qX_rY_s-Y_pX_qY_rY_s),       
        \end{equation}
which only consists of the first four terms in Eq.~\eqref{eq:qeb_op}. The representations of two-body sQEB and QEB operators in the form of configuration basises are shown in Table~\ref{table:1}. It is clear that the sQEB operators are block diagonal in the sense that we can rewrite it as a linear combination of two QEB operators
\begin{equation}
    \tau^{pq}_{rs} = \kappa^{pq}_{rs} + \kappa^{qr}_{sp}
\end{equation}
%It can be implemented using the circuit depicted in Fig.~\ref{fig:sQEB_circuit}, where one control qubit has been removed compared to the QEB circuit in Fig.~\ref{fig:QEB_circuit}. 
The $\tau^{pq}_{rs}$ operator maintains conservation of particle number and the S$_z$ component of the spin if it satisfies the following condition
\begin{equation}
    (\sigma_p+\sigma_q=\sigma_r+\sigma_s) \wedge (\sigma_p+\sigma_s=\sigma_q+\sigma_r)
\end{equation}
When we decompose a two-body sQEB operator into elementary gate sets, it consists of 9 CNOT gates, in contrast to 13 CNOT gates in the QEB circuit (The detailed circuit is provided in the supplementary information).  
%Inspired by Ref.~\citenum{MagEva23b}, one can remove the control-qubit to reduce the CNOT cost, but this may induce symmetry breaking. If one control-qubit is removed, an additional pair of excitations will be included. To ensure that the excitation is a double excitation and satisfies the particle number and total spin projection symmetries, 
%one can incorporate a continuous interchange between $\ket{1001}$ and $\ket{0110}$, thereby yielding the simplified QEB (sQEB) operator. Particle number is obviously conserved, and the total spin projection is perserved 

        \begin{table}[H]
        \centering
        \def\arraystretch{1.5}%  1 is the default, change 
        \begin{tabular}{ |c|c| } 
            \hline
            QEB & sQEB \\ 
            \hline
            $U^{pq}_{rs}\ket{0011} = \cos{\theta}\ket{0011} + \sin{\theta}\ket{1100}$ & $U^{pq}_{rs}\ket{0011} = \cos{\theta}\ket{0011} + \sin{\theta}\ket{1100}$\\ 
            $U^{pq}_{rs}\ket{1100} = \cos{\theta}\ket{1100} - \sin{\theta}\ket{0011}$ & $U^{pq}_{rs}\ket{1100} = \cos{\theta}\ket{1100} - \sin{\theta}\ket{0011}$\\ 
            other unchanged & $U^{pq}_{rs}\ket{1001} = \cos{\theta}\ket{1001} + \sin{\theta}\ket{0110}$\\ 
             & $U^{pq}_{rs}\ket{0110} = \cos{\theta}\ket{0110} - \sin{\theta}\ket{1001}$\\ 
              & other unchanged \\
            \hline
        \end{tabular}
        \caption{Representations of two-body QEB and sQEB operators in the particle number basis $\ket{n_pn_qn_rn_s}$ with $n_q$ being the number of particles that occupy the $q$-th qubit.}
        \label{table:1}
        \end{table}

        Similarly, one can add up two sQEB operators $\tau^{pq}_{rs}$ and $\tau^{qp}_{sr}$ to construct one QEB operator so that 
        \begin{equation}
            e^{\frac{\theta}{2}\tau^{pq}_{rs}} e^{\frac{\theta}{2} \tau^{qp}_{sr}} = e^{\theta \kappa^{pq}_{rs}}.
        \end{equation}
        Hence, the ADAPT-VQE algorithm using the sQEB operator pool can exactly reproduce the results obtained using the QEB operator pool. In addition, one can also extend this technique to the FEB excitation operators by defining simplified-FEB operator as
        \begin{equation}
            (a^\dag_p a^\dag_q a_r  a_s - a^\dag_s a^\dag_r a_q  a_p) + (a^\dag_q a^\dag_r a_s  a_p -a^\dag_p a^\dag_s a_r  a_q), 
        \end{equation} 
        leading to reduction of 4 CNOTs.
        
    \subsection{Excited state approach}
        
        %The one-body QEB excitation and two-body sQEB excitation are collected to form the sQEB operator pool. This pool is then integrated into the ADAPT-VQE framework for solving both ground and excited state problems.

        It is important to utilize quantum computer to find excited states of a many-electron system, which is a fundamental research field in quantum chemistry. There are broadly two types of excited-state quantum algorithms. One is to first determine a reference state, then construct a low-energy subspace by applying the excitation operators onto the reference state, and finally diagonalize the the eigenvalue equation in this subspace \cite{McCKimCar17, McCJiaRub20, OllKanChe20, FanLiuLi21, AstKumAbr23}. Another one is to define an objective function by incorporating specified constraints into the Hamiltonian, and then variationally optimize the parameterized ansatz to minimize the objective function~\cite{NakMitFuj19, HigOscWan19,PerMcCSha14}. For example, a penalty function that enforces orthogonality between the target state and previously determined states can be added to the Hamiltonian to find excited states in the framework of the VQE~\cite{YorBarArv22}. Here, we combine this technique and the sQEB-ADAPT-VQE to tackle excited state problems.

        Once the ground state $\ket{\psi_g}$ is found, one can define an objective function
        \begin{equation}
            H^\prime=H + \alpha\ket{\psi_g}\bra{\psi_g}.
        \end{equation} 
        As long as $\alpha > \Delta E$, where $\Delta E$ is the gap between the ground and first excited state, the global minima of the objective function
        \begin{align}
            E(\theta) &= \bra{\psi_0} U^\dagger(\theta) H^\prime U(\theta) \ket{\psi_0} \\
            &= \bra{\psi_0} U^\dagger(\theta) H U(\theta) \ket{\psi_0} + \alpha {|\bra{\psi_0} U^\dagger(\theta) \ket{\psi_g}|}^2
        \end{align}
        corresponds to the energy of the first excited state. The first term in $E(\theta)$ represents the expectation value of $U(\theta)\ket{\psi_0}$ with respect to $H$, which can be straightforwardly measured. The second term is the overlap between $U(\theta)\ket{\psi_0}$ and $\ket{\psi_g}$, which can be evaluated using the SWAP test or measuring the probability of  $U^\dagger(\theta) \ket{\psi_g}$ collapsing to $\ket{\psi_0}$. In this work, we focus on the calculations of the singlet states, so an additional penalty function is added to the Hamiltonian as
        \begin{equation}
            H^\prime=H + \alpha\ket{\psi_g}\bra{\psi_g} + \beta S^2.
        \end{equation}
        Here, $S^2$ denotes the spin-squared operator, with the singlet state exhibiting the lowest expectation value of zero.
  
        %The choice of the reference state may significantly impact the results obtained using the ADAPT-VQE. However, preparing a reference state that has a substantial overlap with the target excited state is a challenging task. To address this, e-QEB-ADAPT-VQE employs single/full parameter energy reduction as the operator selection criterion, which outperform QEB-ADAPT-VQE.

        Here, we employ the single-parameter energy descent as the criterion of selecting operators, which is demonstrated to be more efficient than the residual gradient scheme in the excited-state calculations~\cite{YorBarArv22}. The single-parameter energy minima is defined as
        \begin{equation}\label{eq:Esp}
            \min_{\theta_\mu} E(\theta_\mu) = \min_{\theta_\mu} \bra{\psi_k} U^\dagger(\theta_\mu) H U(\theta_\mu) \ket{\psi_k}
        \end{equation}
        where $\ket{\psi_k}$ represents the optimized state after the k-th iteration, with energy $E_k = \bra{\psi_k} H \ket{\psi_k}$. The single-parameter energy descent is defined as
        \begin{equation}
            \Delta E_\mu = E_k - \min_{\theta_\mu} E(\theta_\mu)
        \end{equation}
       for each operator in the pool $\{\tau_\mu\}$, and the operator with the largest energy descent will be selected. In Ref.~\citenum{YorBarArv22}, the energy descent is computed via single parameter optimization. Here, we introduce an analytic formalism of the single-parameter energy.
        
       As both QEB and sQEB operators satisfy $\tau^3=-\tau$, one can expand the exponential operator as\cite{CheCheFre21}
        \begin{equation}
            U(\theta)=e^{\theta \tau}=1+\sin{\theta}\tau +(\cos{\theta}-1)(-\tau^2)
        \end{equation}
        As such, the energy functional is rewritten as
        \begin{align*}
            E(\theta)
            &= \bra{\psi_k} U^\dag(\theta) H U(\theta) \ket{\psi_k} \\
            &= \bra{\psi_k} (1 - \sin{\theta}\tau + (\cos{\theta}-1)(-\tau^2)) 
            H
            (1 + \sin{\theta}\tau + (\cos{\theta}-1)(-\tau^2)) \ket{\psi_k} \\
            &= f_0 + f_1\sin{\theta} + f_2\sin{2\theta} + f_3\cos{\theta} + f_4\cos{2\theta}
        \end{align*}
        There are five coefficients $f_0-f_4$ to be measured with respect to $\ket{\psi_k}$ on quantum computers. Their detailed expressions are documented in the supplementary information. After these coefficients are determined, the energy minima can be computed on a classical computer. In contrast to Eq.~\eqref{eq:Esp}, the number of measurements in the sQEB-ADAPT-VQE using the analytical scheme is fixed in the sense that it does not depend on the number of iterations required to optimize the parameter as discussed in Ref.~\citenum{YorBarArv22}.

\section{Numerical results} \label{sec:Results}

    We numerically assess the performance of the ADAPT-VQE algorithm using the sQEB operator pool with respect to the FEB and QEB operator pools. In this section, we apply the new algorithm to study the ground and first singlet excited states of small molecular systems, including LiH, BeH$_2$, H$_6$, with bond lengths ranging from 0.5 to 3.5 \AA. All calculations are performed using the STO-3G basis set, and all Hartree-Fock orbitals were included in the ADAPT-VQE calculations. As such, 12, 14 and 12 qubits are used for LiH, BeH$_2$, H$_6$, respectively.

    We employ the high-performance Q$^2$Chemistry~\cite{FanLiuZen22} simulator to carry out ADAPT-VQE calculations, and PySCF~\cite{SunBerBlu18} to carry out Hartree-Fock calculations and then obtain one- and two-electron integrals. The reference results are computed using the Full Configuration Interaction (FCI) method. OpenFermion~\cite{McCRubSun20} is leveraged to map fermionic operators onto qubit operators. The Broyden-Fletcher-Goldfarb-Shanno algorithm available in Scipy~\cite{VirGomOli20} is employed to minimize the objective function.
    
    %In all the figures depicted below, the data marked by the red, green, and blue colors denote the outcomes obtained from the FEB, QEB, and sQEB operator pools, respectively.

    \subsection{Ground-state calculations}
    
    In case of ground state calculations utilizing three kinds of operator pools, the excitation operators are restricted to promote particles from occupied to virtual orbitals, with $p,q$ and $r,s$ indicating occupied and virtual orbitals. The excitation operators conserving the particle number and $S_z$ symmetry are incorporated into the operator pool. We employ the residual gradients to select operators for updating the wave function ansatz. Initial values of variational parameters corresponding to newly added operators are set to zero. The ADAPT-VQE procedure ends when the 2-norm of the residual gradient vector falls below a predefined threshold $\epsilon$.

    \begin{figure}[!htb]
        \centering
        \makebox[\textwidth][c]{\includegraphics[width=0.99\textwidth]{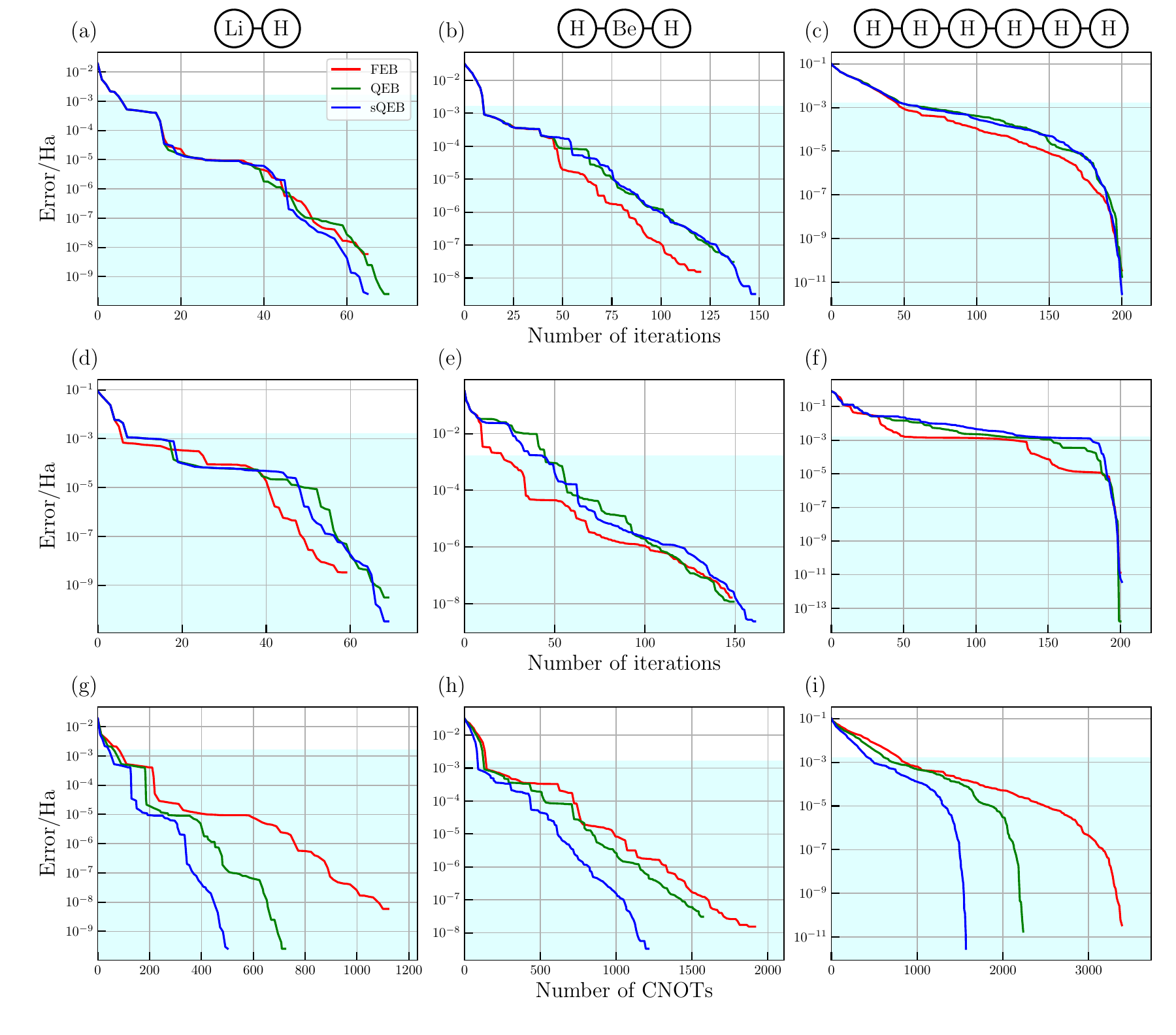}}
        
        \caption{\textbf{Energy convergence curves.} (a-c) Energy errors as a function of the number of iterations for (a) LiH at 1.5 \AA, (b)BeH$_2$ at 1.25 \AA, (c)H$_6$ chain at 1.0 \AA; (d-f) Energy errors as a function of the number of iterations for (d)LiH at 3.0 \AA, (e)BeH$_2$ at 3.0 \AA, (f)H$_6$ chain at 3.0 \AA. (g-i) Energy errors as a function of the number of CNOTs for (g)LiH at 1.5 \AA, (h)BeH$_2$ at 1.25 \AA, (i)H$_6$ chain at 1.0 \AA. Shaded areas indicate the energy errors less than 1.59 millihartree.}
        \label{fig:ground-convergence}
    \end{figure}
    
    \subsubsection{Convergence}\label{sec:conv}
    Figure~\ref{fig:ground-convergence} illustrates the convergence curves of the ADAPT-VQE algorithms using FEB, QEB and sQEB operator pools for three molecules, including LiH, BeH$_2$, and H$_6$ chain with each hydrogen atom equispaced along a line. Figure~\ref{fig:ground-convergence} (a)-(c) present the convergence behavior of the total energy errors as a function of the number of iterations for three molecules near the equilibrium bond lengths. In case of these weakly correlated systems, the ADAPT-VQE using all three kinds of operator pools exhibits similar convergence behavior. In case of  LiH at 1.5 \AA, insignificant deviations are only observed after the energy errors are less than $\sim 10^{-5}$ Hartree. In case of BeH$_2$ at 1.25 \AA, the energy errors of the ADAPT-VQE using all three kinds of operator pools also decrease very fast to 1 millihartree after $\sim$10 iterations. In contrast, the convergence of the ground-state energy for H$_6$ is much slower than LiH and BeH$_2$. 
    
    %discernible differences emerge until the error decrease below chemical accuracy, with the FEB requiring approximately 20 fewer steps to achieve the same level of accuracy compared to the QEB and sQEB pools. The convergence curve for the latter two pools exhibit similar behavior, intertwining with each other. For the H$_6$ chain, the FEB pool exhibits approximately $25\%$ faster convergence, while the convergence curves for the sQEB and QEB pools nearly coincide.
    
    Figure~\ref{fig:ground-convergence} (d)-(f) show the energy convergence curves as a function of the number of iterations for three molecule with larger bond length, that is atoms in these molecules are well separated. In case of all three molecules, the ADAPT-VQE algorithm using the FEB operator pool exhibits faster convergence rate to chemical accuracy (1.59 millihartree). Especially for the BeH$_2$ molecule at 3.0 \AA, the energy error of the FEB-ADAPT-VQE approach rapidly falls below 1 millihartree after $\sim$25 iteration, whereas both QEB-ADAPT-VQE and sQEB-ADAPT-VQE encounter plateau, leading to almost doubling the number of iterations in order to achieve the same accuracy. In the case of the strongly correlated H$_6$ system, the ADAPT-VQE algorithm using three kinds of operator pools struggles to achieve chemical accuracy, exhibiting prolonged plateaus in the convergence curves when the energy errors are larger than 1 millihartree. For LiH at 3.0 \AA, the convergence curves of the QEB-ADAPT-VQE and sQEB-ADAPT-VQE also exhibit obvious plateaus when the energy errors approach 1 millihartree. Such a kind of plateaus also appear in the energy convergence curves of LiH at 1.5 \AA, when the energy errors are $\sim6 \times 10^{-4}$ and $\sim1 \times 10^{-5}$ Hartree.
    
    Although the FEB-ADAPT-VQE requires fewer iterations, namely fewer FEB unitary operations, to achieve the same accuracy as the QEB-ADAPT-VQE and sQEB-ADAPT-VQE, implementing these FEB operations on a quantum computer requires a larger number of CNOTs in contrast to the QEB and sQEB operations. Figure~\ref{fig:ground-convergence} (g)-(i) depict the energy error curves as a function of the number of CNOTs for three molecules near the equilibrium bond lengths. It is clear that, in order to achieve the same level of accuracy, the sQEB-ADAPT-VQE typically requires fewer number of CNOTs than both QEB-ADAPT and FEB-ADAPT-VQE. As the energy errors decrease, the difference of the number of CNOTs between the sQEB-ADAPT-VQE algorithm and other two algorithms becomes more apparent. In addition, the QEB-ADAPT-VQE generally demands fewer number of CNOTs than the FEB-ADAPT-VQE. This conclusion is consistent with results presented in  Ref.~\citenum{YorBarArv22}.

    \subsubsection{Potential energy curves} 
\begin{figure}[!htb]
        \centering
        \makebox[\textwidth][c]{\includegraphics[width=0.99\textwidth]{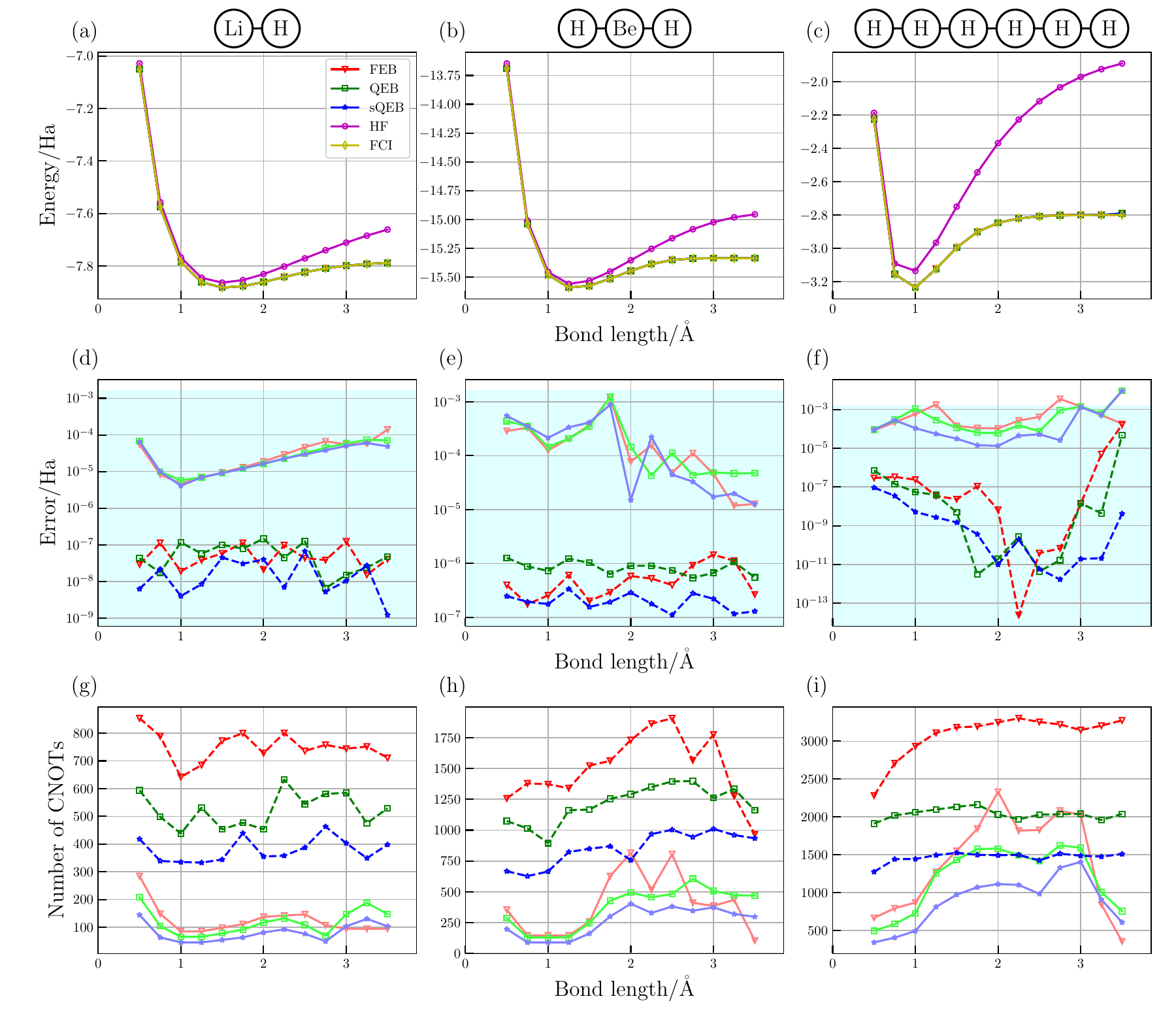}}
        \caption{\textbf{Ground-state Potential Energy Surfaces (PESs).} (a)-(c) PESs of LiH, BeH$_2$ and H$_6$ molecules, computed by the ADAPT-VQE using FEB, QEB and sQEB operator pools, and Hartree-Fock and FCI methods; (d)-(f) Energy errors when the convergence criteria is set to $\epsilon=10^{-3}$ (light red for FEB, light green for QEB, light blue for sQEB), and $\epsilon=10^{-5}$ (deep red for FEB, deep green for QEB, deep blue for sQEB); (g)-(i) The number of CONTs needed to achieve certain accuracy of $10^{-3}$ (light color) and $10^{-6}$ Hartree (deep color) for three pools.}
        \label{fig:ground-PES}
    \end{figure}

    Figure~\ref{fig:ground-PES}(a)-(c) shows Potential Energy Surfaces (PESs) computed by the ADAPT-VQE using FEB, QEB and sQEB  operator pools. The results computed by the Hartree-Fock and FCI methods are also shown for comparison. The bond lengths of Li-H, Be-H and H-H in three molecules vary from 0.5 to 3.5 \AA. It is obvious that, as the geometry structures of three molecules transits from equilibrium to dissociated ones, the correlation effect becomes much stronger so that Hartree-Fock fails to recover the exact ground-state energies. Especially for the H$_6$ molecule, the energy error is as large as $\sim$0.9 Hartree at $R_{H-H}=3.5$ \AA. All three kinds of ADAPT-VQE schemes are able to accurately reproduce the FCI results when a tight convergence criteria $\epsilon=10^{-5}$ is used as discussed in the following. In the work, the convergence criteria is much tighter than that used in the original ADAPT-VQE work, in which the tightest one is $\epsilon=10^{-3}$. This difference mainly results from different kinds of excitation operators used to generate the operator pools. As discussed in Ref.~\citenum{LiuWanLi20}, the operator pools consisting of general single and double excitations are more stable because the corresponding convergence criteria is related to the anti-Hermitian contracted Schr\"odinger equation.
    %Panels   in Figure ~\ref{fig:ground-PES} display the energies obtained from the three operator pools, compared with the standard Hartree-Fock and FCI results. The energies calculated by ADAPT-VQE using the FEB, QEB and sQEB operator pools Coincide with the FCI energy curve.

    Figure~\ref{fig:ground-PES} (d)-(f) show the energy errors with respect to the FCI results for three molecules at different bond lengths. The calculations were performed with the convergence threshold $\epsilon$ set to $10^{-3}$ and $10^{-5}$. The overall performance of FEB-ADAPT-VQE, QEB-ADAPT-VQE and sQEB-ADAPT-VQE is similar. When $\epsilon = 10^{-3}$, all three ADAPT-VQE algorithms are able to achieve chemical accuracy for LiH and BeH$_2$ across all bond lengths, while all of them fail to converge to chemical accuracy over a wide range of H-H bond lengths for the H$_6$ chain. When $\epsilon$ is set to $10^{-5}$, FEB-ADAPT-VQE, QEB-ADAPT-VQE and sQEB-ADAPT-VQE are able to converge to a very high accuracy for LiH,  BeH$_2$ and H$_6$. In contrast to FEB-ADAPT-VQE and QEB-ADAPT-VQE, sQEB-ADAPT-VQE exhibits slightly better performance with lower energy errors. For example, in case of H$_6$, the energy errors of FEB-ADAPT-VQE and QEB-ADAPT-VQE are $\sim 10^{-4}$ Hartree while sQEB-ADAPT-VQE has an energy error of less than $10^{-8}$ Hartree at a large H-H bond length of 3.5 \AA. 
    %all pools converge to high accuracy levels, except at large bond lengths for the H$_6$ chain due to its strongly correlated nature. These finding suggest that the sQEB operator pool can effectively construct ansatz that systematically approximate the ground state for these molecules within the ADAPT-VQE framework.

    Given an energy error, the number of CNOTs required to implement the ADAPT-VQE algorithms for predicting the potential energy curves using three kinds of operator pools are shown in Figure~\ref{fig:ground-PES} (g)-(i). As discussed in section~\ref{sec:conv}, the sQEB-ADAPT-VQE algorithm requires fewer number of CNOTs to achieve the same accuracy as FEB-ADAPT-VQE and QEB-ADAPT-VQE for small molecules with nearly equilibrium bond lengths. As shown in Figure~\ref{fig:ground-PES} (g)-(i), this conclusion is still maintained in the ADAPT-VQE calculations of these molecules with structures ranging from equilibrium to dissociated states using a tight energy error of $10^{-6}$ Hartree. When using a loose energy error $10^{-3}$ Hartree, the FEB-ADAPT-VQE may produce shallower circuits than the sQEB-ADAPT-VQE for some dissociated molecular structures. In contrast to the QEB-ADAPT-VQE, the sQEB-ADAPT-VQE is able to reduce the number of CNOTs by $\sim28\%$ as illustrated in Table~\ref{table:2}. Here, 
    the ratios of reduced CNOT count are estimated by $1-n_{sQEB}/n_{QEB}$, where $n_{sQEB}$ and $n_{QEB}$ are the number of CNOTs required in the sQEB-ADAPT-VQE and QEB-ADAPT-VQE calculations, respectively, and we average these ratios over all bond lengths. Table~\ref{table:2} reveals that sQEB requires approximately $28\%$ fewer CNOT count than QEB to achieve the same level of accuracy, which is quite consistent with the ratio of reduction in the number of CNOTs required for implementing two-qubit sQEB and QEB gates, which is $1-\frac{9}{13} \approx 31\%$.
    
    %The advantage of employing the sQEB excitation lies in its reduced requirement for CNOTs compared to the QEB excitation. Figure ~\ref{fig:ground-PES} (g)-(i) illustrates the number of CNOTs needed to achieve errors of $10^{-3}$ and $10^{-6}$ Hartree for the three operator pools. For achieve chemical accuracy (error of $10^{-3}$ Hartree), the sQEB pool consistently requires fewer CNOTs than the QEB pool, while the FEB pool necessitates the most CNOTs, except at large bond lengths. When aiming for an error of $10^{-6}$ Hartree, the sQEB pool exhibits the lowest CNOT count, while the QEB pool requiring fewer CNOTs than the FEB pool in most instance. Therefore, the sQEB operator pool emerges as the most CNOT-efficient choice for calculately computing the potential energy surface of the three molecules.

    \begin{table}[H]
    \centering
    \def\arraystretch{1.5}%  1 is the default, change 
    \begin{tabular}{ c|c c c|c c } 
        \hline
        \multirow{2}{*}{} & \multicolumn{3}{c|}{Groud state} & \multicolumn{2}{c}{Excited state} \\ 
        & LiH & BeH$_2$ & H$_6$ chain & LiH & BeH$_2$ \\ 
        \hline
        Low accuracy & 31.05 & 30.19 & 26.07 & 32.62 & 25.86 \\ 
        High accuracy & 27.00 & 29.73 & 27.95 & 27.68 & 21.70 \\ 
        \hline
    \end{tabular}
    \caption{Ratios of reduction in the number of CNOTs required in the sQEB-ADAPT-VQE calculations to the number required in the QEB-ADAPT-VQE calculations. A low-accuracy threshold ($10^{-3}$ Hartree) and a high-accuracy threshold ($10^{-6}$ Hartree) are used to perform the ADAPT-VQE calculations.}
    \label{table:2}
    \end{table}

    \subsection{Excited-state calculations}

    We further apply the sQEB-ADAPT-VQE to simulate electronically excited states of LiH and BeH$_2$. Generalized excitation operators are utilized to generate the operator pools. We adopt the single-parameter energy reduction $\Delta E$ as the criteria for selecting operators. The value from single-parameter optimization is used as the initial value for the newly added parameter in the subsequent VQE optimization. The convergence of the ADAPT-VQE procedure is achieved when the 2-norm of the $\Delta E$ vector is less than a predefined threshold $\epsilon$. Since we focus on singlet states in this work, a penalty term associated with the total spin operator, $S^2$, is incorporated into the cost function. As mentioned in Ref.~\citenum{YorBarArv22}, the e-QEB-ADAPT-VQE failed to achieve chemical accuracy in simulating the first singlet excited state under certain geometry if the Hartree-Fock state is used as the initial state. In this work, we choose a single configuration state function (CSF) as the initial state, which can be determined from low-cost classical methods, such as configuration interaction singles and doubles (CISD). The ground state is extracted from the ADAPT-VQE calculations with a higher convergence threshold.

\begin{figure}[H]
        \centering
        \makebox[\textwidth][c]{\includegraphics[width=0.33\textwidth]{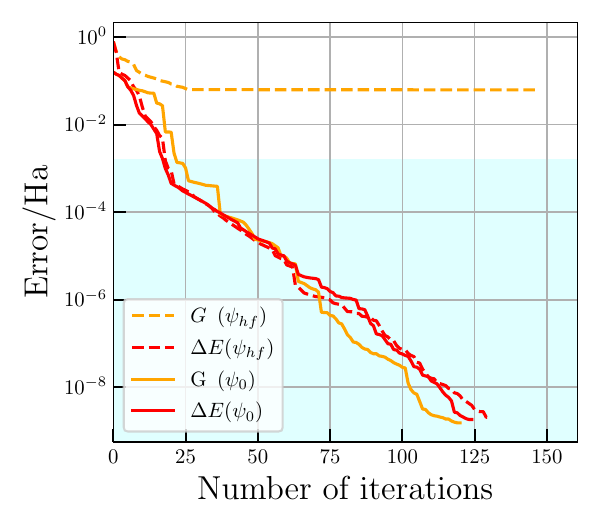}}
        
        \caption{Energy convergence curves of the FEB-ADAPT-VQE for the first singlet excited state of BeH$_2$ at 2.0 \AA. Gradient-based (orange line) and $\Delta$E-based (red line) operator selection criterion are shown, and the Hartree-Fock state $\ket{\psi_{hf}}$ (dotted line) and the configuration state function $\ket{\psi_0}$ (solid line) are considered the initial state.}
        \label{fig:g_vs_e}
    \end{figure}

    Figure ~\ref{fig:g_vs_e} compares the convergence of the FEB-ADAPT-VQE using gradient- and $\Delta$E-based criteria for selecting operators. The target state is the first singlet excited state of the BeH$_2$ molecule at a bond length of 2.0 \AA. The Hartree-Fock state and the single CSF that has the largest coefficient in the classical CISD calculations are employed as the initial state. When the Hartree-Fock state is used, the ADAPT-VQE using the gradient-based criteria fails to find the correct state, whereas the ADAPT-VQE using the $\Delta$E-based criteria succeeds. When a more appropriate initial state is used, the ADAPT-VQE using both gradient-based and $\Delta$E-based criteria successfully converges to the correct state. %The $\Delta$E selection criterion requires fewer steps to achieve chemical accuracy, demonstrating its superiority.
    
\begin{figure}[H]
        \centering
        \makebox[\textwidth][c]{\includegraphics[width=0.99\textwidth]{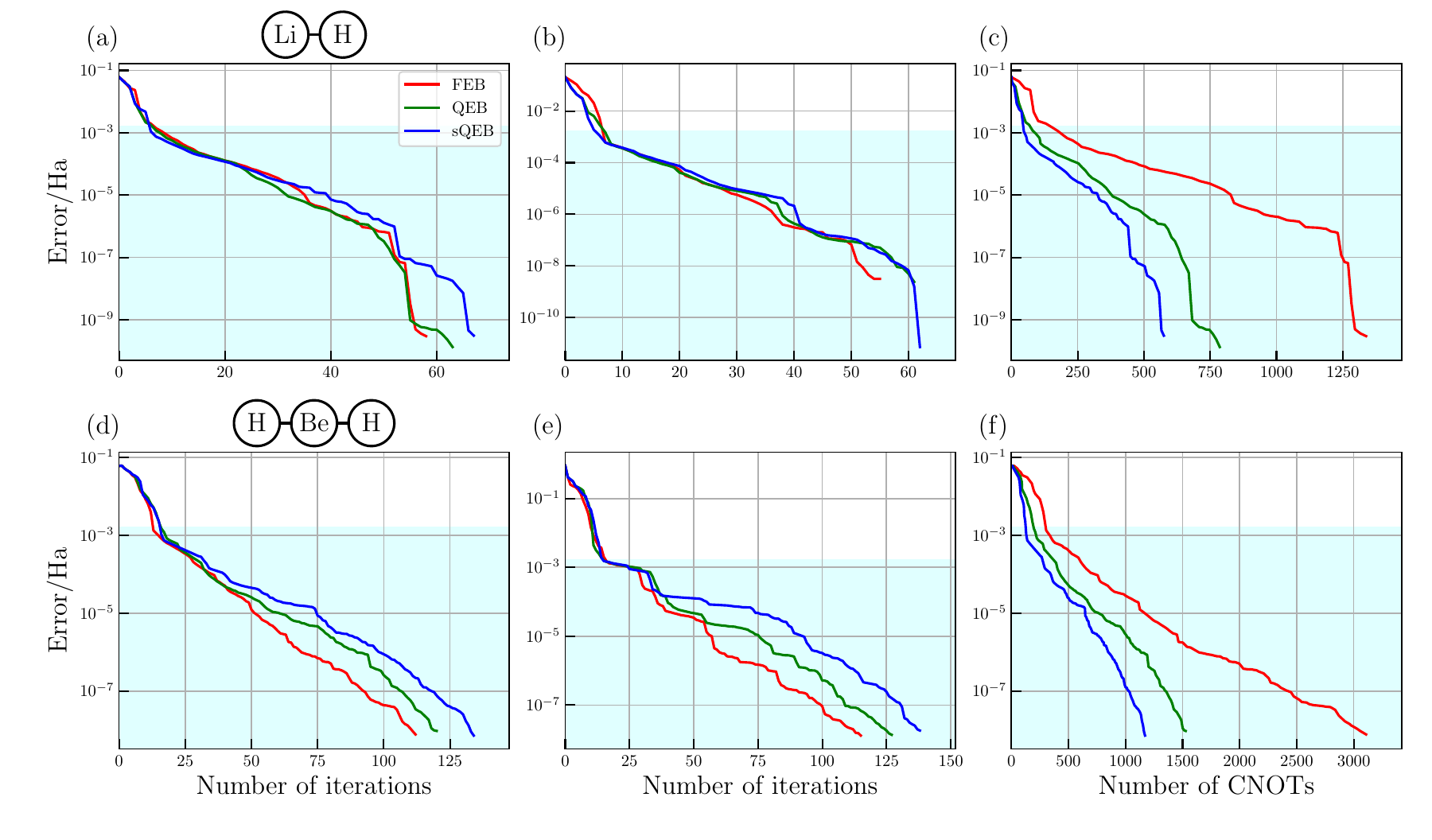}}
        
        \caption{\textbf{Energy Convergence} (a) and (d) depict energy errors as a function of the number of iterations for LiH at 1.5 \AA and BeH$_2$ at 1.25 \AA, respectively. (b) and (e) energy errors as a function of the number of iterations for LiH at 3.0 \AA and BeH$_2$ at 3.0 \AA, respectively. (c) and (f) depicts energy errors as a function of the number of CNOTs for LiH at 1.5 \AA, and BeH$_2$ at 1.25 \AA. The shaded area indicates the energy error less than 1 kcal/mol.}
        \label{fig:excited-convergence}
    \end{figure}

    Figure ~\ref{fig:excited-convergence} illustrates the energy convergence curves for LiH and BeH$_2$ molecules. Here, Figure~\ref{fig:excited-convergence} (a) and (d) depict the convergence curves near the equilibrium bond length, and (b) and (e) depict the convergence curves at the significantly stretched bond length. The ADAPT-VQE using three kinds of operator pools exhibits similar convergence behaviors for LiH, while it shows slight difference for BeH$_2$ when the energy errors fall below 1 millihartree. Like the ground-state simulations, the FEB-ADAPT-VQE exhibits faster convergence than the QEB-ADAPT-VQE and sQEB-ADAPT-VQE in the excited-state simulations. Figure ~\ref{fig:excited-convergence} (c) and (f) show the energy convergence curves with respect to the CNOT count. Despite the sQEB-ADAPT-VQE necessitates a larger number of iterations to achieve the same accuracy, it requires fewer CNOTs than the FEB-ADAPT-VQE and QEB-ADAPT-VQE, demonstrating its advantage in terms of CNOT-efficiency.
    %whereas the sQEB-ADAPT-VQE necessitates a larger number of iterations compared to QEB pool.
    
    %Figure ~\ref{fig:excited-convergence} (b) and (e) delineate convergence profiles at long bond lengths. For LiH molecule, three pools have similarly behavior, the curve nearly coincided until exhibit analogous behaviors. For BeH$_2$, the distinctions emerge only within chemical accuracy thresholds, FEB pool exhibiting superior performance and sQEB requiring a greater number of iterations to achieve the same accuracy. Figure ~\ref{fig:excited-convergence} (c) and (f) present energy convergence curve with respect to CNOT count. Despite sQEB pool necessitating a greater number of iterations to achieve the same accuracy, it employs fewer CNOTs than other two pools, underscoring its advantage in terms of CNOT-efficiency.

    In the ADAPT-VQE, determining the operator used to update the wave function ansatz requires
    measurement of residual gradients or single-parameter energy reductions $\Delta E$ for all operators within the pool. Here, we discuss the measurement costs associated with this subroutine for the ADAPT-VQE using three kinds of operator pools. For each operator $\tau$, its corresponding residual gradient is given by $\bra{\psi_k} [H,\tau] \ket{\psi_k}$, namely the expectation value of the commutator $[H,\tau]$ with respect to the state $\ket{\psi_k}$. To measure this quantity, each Pauli string within the commutator, denoted as $M_\tau=\{P_i\mid P_i \in\ [H,\tau]\}$, should be measured. The collection of all Pauli strings, denoted as $M=\bigcup\limits_\tau M_\tau$, represents the set of Pauli strings that need to be measured. If the single-parameter energy reduction is employed as the criteria for operator selection, additional measurements are necessary (detailed in supporting information). 
    %The collection of all Pauli strings within these operators represnts the total set of Pauli strings that need to be measured.

    \begin{figure}[H]
        \centering
        \makebox[\textwidth][c]{\includegraphics[width=0.5\textwidth]{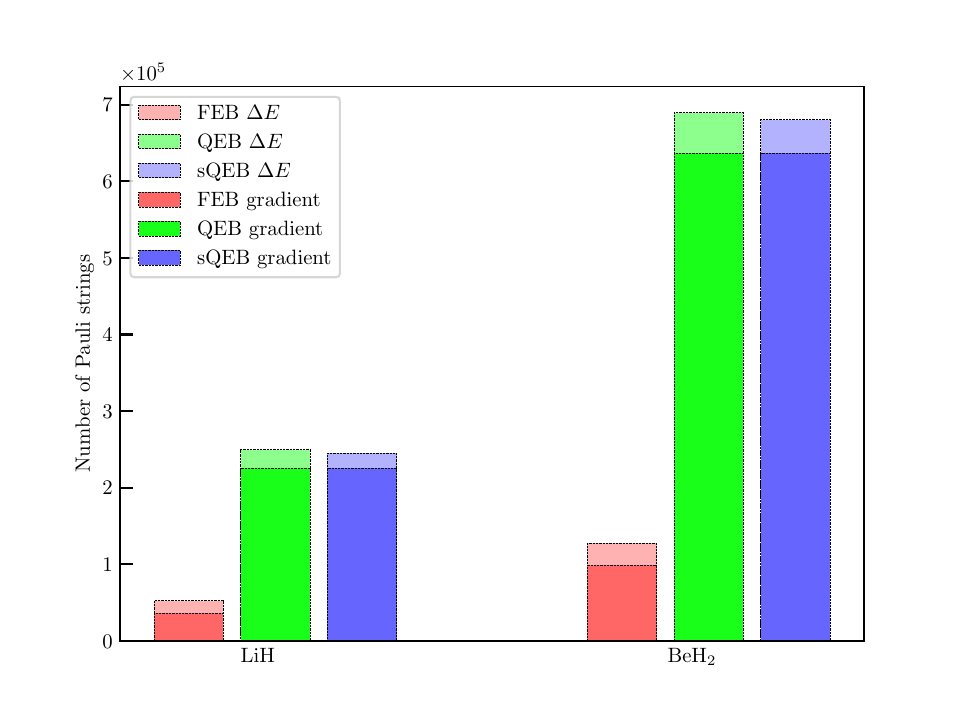}}
        \caption{\textbf{Number of Pauli strings need to be measured for the FEB-ADAPT-VQE, QEB-ADAPT-VQE and sQEB-ADAPT-VQE.}}
        \label{fig:Paulistrs}
    \end{figure}
    
    We collect the number of Pauli strings that are required to be measured when selecting operators based on residual gradients and $\Delta E$ for three kinds of operator pools generated from generalized one- and two-body excitation operators, and the results are depicted in Fig.~\ref{fig:Paulistrs}. It is evident that the number of Pauli strings required to be measured for the sQEB-ADAPT-VQE and QEB-ADAPT-VQE calculations of LiH and BeH$_2$ is nearly identical, whereas the FEB-ADAPT-VQE necessitates significantly fewer Pauli strings. The additional overhead in measurement incurred by the $\Delta E$-based criteria remains acceptable in comparison with the gradient-based criteria. Thus, applying the analytical method introduced in the work for calculating $\Delta E$ is highly resource efficient. This provides us a compelling alternative to the gradient-based criteria.

\section{Conclusion and outlook} \label{sec:Conclusion}
    
    In this study, we introduce simplified qubit excitation-based operators to build the wave function ansatz within the framework of ADAPT-VQE. In contrast to the QEB operators, the sQEB operators can be implemented using a quantum circuit with only 9 CNOTs, leading to 4 CNOT reduction. Numerically, we assess the performance of the ADAPT-VQE using the sQEB operator pool, and compare it against the QEB-ADAPT-VQE and FEB-ADAPT-VQE algorithms. In case of three small molecules, including LiH, BeH$_2$, and H$_6$ chain, we observe that the sQEB operator pool is able to achieve high accuracy convergence levels, comparable to both QEB and FEB operator pools. Notably, the FEB-ADAPT-VQE demonstrates its superiority in convergence at a specified accuracy, while the QEB-ADAPT-VQE and sQEB-ADAPT-VQE exhibit an advantage in building shallow-circuit Ans\"atze. On average, the sQEB-ADAPT-VQE necessitates $28\%$ fewer CNOT operations than the QEB-ADAPT-VQE in ground-state calculations. In case of excited-state calculations, the sQEB-ADAPT-VQE exhibits a very similar performance. We believe that the sQEB-ADAPT-VQE algorithm provides us a promising scheme to carry out quantum simulations of electroni structure on near-term quantum devices with limited circuit depths. 

\section{Acknowledgments}
This work is supported by Innovation Program for Quantum Science and Technology (2021ZD0303306), by the National Natural Science Foundation of China (22073086, 22288201),  Anhui Initiative in Quantum Information Technologies (AHY090400).

\footnotesize{
    \bibliography{qc}

\providecommand{\latin}[1]{#1}
\makeatletter
\providecommand{\doi}
  {\begingroup\let\do\@makeother\dospecials
  \catcode`\{=1 \catcode`\}=2 \doi@aux}
\providecommand{\doi@aux}[1]{\endgroup\texttt{#1}}
\makeatother
\providecommand*\mcitethebibliography{\thebibliography}
\csname @ifundefined\endcsname{endmcitethebibliography}  {\let\endmcitethebibliography\endthebibliography}{}
\begin{mcitethebibliography}{60}
\providecommand*\natexlab[1]{#1}
\providecommand*\mciteSetBstSublistMode[1]{}
\providecommand*\mciteSetBstMaxWidthForm[2]{}
\providecommand*\mciteBstWouldAddEndPuncttrue
  {\def\EndOfBibitem{\unskip.}}
\providecommand*\mciteBstWouldAddEndPunctfalse
  {\let\EndOfBibitem\relax}
\providecommand*\mciteSetBstMidEndSepPunct[3]{}
\providecommand*\mciteSetBstSublistLabelBeginEnd[3]{}
\providecommand*\EndOfBibitem{}
\mciteSetBstSublistMode{f}
\mciteSetBstMaxWidthForm{subitem}{(\alph{mcitesubitemcount})}
\mciteSetBstSublistLabelBeginEnd
  {\mcitemaxwidthsubitemform\space}
  {\relax}
  {\relax}

\bibitem[Dirac(1929)]{Dirac29}
Dirac,~P. A.~M. Quantum mechanics of many-electron systems. \emph{Proc.\ R.\ Soc.\ A} \textbf{1929}, \emph{123}, 714--733\relax
\mciteBstWouldAddEndPuncttrue
\mciteSetBstMidEndSepPunct{\mcitedefaultmidpunct}
{\mcitedefaultendpunct}{\mcitedefaultseppunct}\relax
\EndOfBibitem
\bibitem[Feynman \latin{et~al.}(1982)Feynman, \latin{et~al.} others]{Fey82}
Feynman,~R.~P.; others Simulating physics with computers. \emph{Int.\ J.\ Theor.\ Phys.} \textbf{1982}, \emph{21}\relax
\mciteBstWouldAddEndPuncttrue
\mciteSetBstMidEndSepPunct{\mcitedefaultmidpunct}
{\mcitedefaultendpunct}{\mcitedefaultseppunct}\relax
\EndOfBibitem
\bibitem[Cao \latin{et~al.}(2019)Cao, Romero, Olson, Degroote, Johnson, Kieferov{\'a}, Kivlichan, Menke, Peropadre, Sawaya, \latin{et~al.} others]{CaoYudRom19}
Cao,~Y.; Romero,~J.; Olson,~J.~P.; Degroote,~M.; Johnson,~P.~D.; Kieferov{\'a},~M.; Kivlichan,~I.~D.; Menke,~T.; Peropadre,~B.; Sawaya,~N.~P.; others Quantum chemistry in the age of quantum computing. \emph{Chem.\ Rev.} \textbf{2019}, \emph{119}, 10856--10915\relax
\mciteBstWouldAddEndPuncttrue
\mciteSetBstMidEndSepPunct{\mcitedefaultmidpunct}
{\mcitedefaultendpunct}{\mcitedefaultseppunct}\relax
\EndOfBibitem
\bibitem[McArdle \latin{et~al.}(2020)McArdle, Endo, Aspuru-Guzik, Benjamin, and Yuan]{McASamEnd20}
McArdle,~S.; Endo,~S.; Aspuru-Guzik,~A.; Benjamin,~S.~C.; Yuan,~X. Quantum computational chemistry. \emph{Rev.\ Mod.\ Phys.} \textbf{2020}, \emph{92}, 015003\relax
\mciteBstWouldAddEndPuncttrue
\mciteSetBstMidEndSepPunct{\mcitedefaultmidpunct}
{\mcitedefaultendpunct}{\mcitedefaultseppunct}\relax
\EndOfBibitem
\bibitem[Kitaev(1995)]{kitaev95}
Kitaev,~A.~Y. Quantum measurements and the Abelian stabilizer problem. \emph{arXiv:quant-ph/9511026} \textbf{1995}, \relax
\mciteBstWouldAddEndPunctfalse
\mciteSetBstMidEndSepPunct{\mcitedefaultmidpunct}
{}{\mcitedefaultseppunct}\relax
\EndOfBibitem
\bibitem[Abrams and Lloyd(1999)Abrams, and Lloyd]{abrams99}
Abrams,~D.~S.; Lloyd,~S. Quantum algorithm providing exponential speed increase for finding eigenvalues and eigenvectors. \emph{Phys.\ Rev.\ Lett.} \textbf{1999}, \emph{83}, 5162\relax
\mciteBstWouldAddEndPuncttrue
\mciteSetBstMidEndSepPunct{\mcitedefaultmidpunct}
{\mcitedefaultendpunct}{\mcitedefaultseppunct}\relax
\EndOfBibitem
\bibitem[Aspuru-Guzik \latin{et~al.}(2005)Aspuru-Guzik, Dutoi, Love, and Head-Gordon]{AspDutLov05}
Aspuru-Guzik,~A.; Dutoi,~A.~D.; Love,~P.~J.; Head-Gordon,~M. Simulated Quantum Computation of Molecular Energies. \emph{Science} \textbf{2005}, \emph{309}, 1704\relax
\mciteBstWouldAddEndPuncttrue
\mciteSetBstMidEndSepPunct{\mcitedefaultmidpunct}
{\mcitedefaultendpunct}{\mcitedefaultseppunct}\relax
\EndOfBibitem
\bibitem[Preskill(2018)]{Pre18}
Preskill,~J. Quantum computing in the NISQ era and beyond. \emph{Quantum} \textbf{2018}, \emph{2}, 79\relax
\mciteBstWouldAddEndPuncttrue
\mciteSetBstMidEndSepPunct{\mcitedefaultmidpunct}
{\mcitedefaultendpunct}{\mcitedefaultseppunct}\relax
\EndOfBibitem
\bibitem[Bharti \latin{et~al.}(2022)Bharti, Cervera-Lierta, Kyaw, Haug, Alperin-Lea, Anand, Degroote, Heimonen, Kottmann, Menke, \latin{et~al.} others]{BhaCerKya22}
Bharti,~K.; Cervera-Lierta,~A.; Kyaw,~T.~H.; Haug,~T.; Alperin-Lea,~S.; Anand,~A.; Degroote,~M.; Heimonen,~H.; Kottmann,~J.~S.; Menke,~T.; others Noisy intermediate-scale quantum algorithms. \emph{Rev.\ Mod.\ Phys.} \textbf{2022}, \emph{94}, 015004\relax
\mciteBstWouldAddEndPuncttrue
\mciteSetBstMidEndSepPunct{\mcitedefaultmidpunct}
{\mcitedefaultendpunct}{\mcitedefaultseppunct}\relax
\EndOfBibitem
\bibitem[Peruzzo \latin{et~al.}(2014)Peruzzo, McClean, Shadbolt, Yung, Zhou, Love, Aspuru-Guzik, and O’brien]{PerMcCSha14}
Peruzzo,~A.; McClean,~J.; Shadbolt,~P.; Yung,~M.-H.; Zhou,~X.-Q.; Love,~P.~J.; Aspuru-Guzik,~A.; O’brien,~J.~L. A variational eigenvalue solver on a photonic quantum processor. \emph{Nat.\ Chem.} \textbf{2014}, \emph{5}, 4213\relax
\mciteBstWouldAddEndPuncttrue
\mciteSetBstMidEndSepPunct{\mcitedefaultmidpunct}
{\mcitedefaultendpunct}{\mcitedefaultseppunct}\relax
\EndOfBibitem
\bibitem[McClean \latin{et~al.}(2016)McClean, Romero, Babbush, and Aspuru-Guzik]{McCJarRom16}
McClean,~J.~R.; Romero,~J.; Babbush,~R.; Aspuru-Guzik,~A. The theory of variational hybrid quantum-classical algorithms. \emph{New\ J.\ Phys.} \textbf{2016}, \emph{18}, 023023\relax
\mciteBstWouldAddEndPuncttrue
\mciteSetBstMidEndSepPunct{\mcitedefaultmidpunct}
{\mcitedefaultendpunct}{\mcitedefaultseppunct}\relax
\EndOfBibitem
\bibitem[Wang \latin{et~al.}(2019)Wang, Higgott, and Brierley]{WanHigBri19}
Wang,~D.; Higgott,~O.; Brierley,~S. Accelerated variational quantum eigensolver. \emph{Phys.\ Rev.\ Lett.} \textbf{2019}, \emph{122}, 140504\relax
\mciteBstWouldAddEndPuncttrue
\mciteSetBstMidEndSepPunct{\mcitedefaultmidpunct}
{\mcitedefaultendpunct}{\mcitedefaultseppunct}\relax
\EndOfBibitem
\bibitem[Hempel \latin{et~al.}(2018)Hempel, Maier, Romero, McClean, Monz, Shen, Jurcevic, Lanyon, Love, Babbush, \latin{et~al.} others]{HemMaiRom18}
Hempel,~C.; Maier,~C.; Romero,~J.; McClean,~J.; Monz,~T.; Shen,~H.; Jurcevic,~P.; Lanyon,~B.~P.; Love,~P.; Babbush,~R.; others Quantum chemistry calculations on a trapped-ion quantum simulator. \emph{Phys.\ Rev.~X} \textbf{2018}, \emph{8}, 031022\relax
\mciteBstWouldAddEndPuncttrue
\mciteSetBstMidEndSepPunct{\mcitedefaultmidpunct}
{\mcitedefaultendpunct}{\mcitedefaultseppunct}\relax
\EndOfBibitem
\bibitem[Nam \latin{et~al.}(2020)Nam, Chen, Pisenti, Wright, Delaney, Maslov, Brown, Allen, Amini, Apisdorf, \latin{et~al.} others]{NamChePis20}
Nam,~Y.; Chen,~J.-S.; Pisenti,~N.~C.; Wright,~K.; Delaney,~C.; Maslov,~D.; Brown,~K.~R.; Allen,~S.; Amini,~J.~M.; Apisdorf,~J.; others Ground-state energy estimation of the water molecule on a trapped-ion quantum computer. \emph{Npj\ Quantum\ Inf.} \textbf{2020}, \emph{6}, 33\relax
\mciteBstWouldAddEndPuncttrue
\mciteSetBstMidEndSepPunct{\mcitedefaultmidpunct}
{\mcitedefaultendpunct}{\mcitedefaultseppunct}\relax
\EndOfBibitem
\bibitem[Shen \latin{et~al.}(2017)Shen, Zhang, Zhang, Zhang, Yung, and Kim]{SheZhaZha17}
Shen,~Y.; Zhang,~X.; Zhang,~S.; Zhang,~J.-N.; Yung,~M.-H.; Kim,~K. Quantum implementation of the unitary coupled cluster for simulating molecular electronic structure. \emph{Phys.\ Rev.~A} \textbf{2017}, \emph{95}, 020501\relax
\mciteBstWouldAddEndPuncttrue
\mciteSetBstMidEndSepPunct{\mcitedefaultmidpunct}
{\mcitedefaultendpunct}{\mcitedefaultseppunct}\relax
\EndOfBibitem
\bibitem[O’Malley \latin{et~al.}(2016)O’Malley, Babbush, Kivlichan, Romero, McClean, Barends, Kelly, Roushan, Tranter, Ding, \latin{et~al.} others]{OMalley16}
O’Malley,~P.~J.; Babbush,~R.; Kivlichan,~I.~D.; Romero,~J.; McClean,~J.~R.; Barends,~R.; Kelly,~J.; Roushan,~P.; Tranter,~A.; Ding,~N.; others Scalable quantum simulation of molecular energies. \emph{Phys.\ Rev.~X} \textbf{2016}, \emph{6}, 031007\relax
\mciteBstWouldAddEndPuncttrue
\mciteSetBstMidEndSepPunct{\mcitedefaultmidpunct}
{\mcitedefaultendpunct}{\mcitedefaultseppunct}\relax
\EndOfBibitem
\bibitem[Colless \latin{et~al.}(2018)Colless, Ramasesh, Dahlen, Blok, Kimchi-Schwartz, McClean, Carter, de~Jong, and Siddiqi]{ColRamDah18}
Colless,~J.~I.; Ramasesh,~V.~V.; Dahlen,~D.; Blok,~M.~S.; Kimchi-Schwartz,~M.~E.; McClean,~J.~R.; Carter,~J.; de~Jong,~W.~A.; Siddiqi,~I. Computation of molecular spectra on a quantum processor with an error-resilient algorithm. \emph{Phys.\ Rev.~X} \textbf{2018}, \emph{8}, 011021\relax
\mciteBstWouldAddEndPuncttrue
\mciteSetBstMidEndSepPunct{\mcitedefaultmidpunct}
{\mcitedefaultendpunct}{\mcitedefaultseppunct}\relax
\EndOfBibitem
\bibitem[Tilly \latin{et~al.}(2022)Tilly, Chen, Cao, Picozzi, Setia, Li, Grant, Wossnig, Rungger, Booth, \latin{et~al.} others]{TilJulChe22}
Tilly,~J.; Chen,~H.; Cao,~S.; Picozzi,~D.; Setia,~K.; Li,~Y.; Grant,~E.; Wossnig,~L.; Rungger,~I.; Booth,~G.~H.; others The variational quantum eigensolver: a review of methods and best practices. \emph{Phys.\ Rep.} \textbf{2022}, \emph{986}, 1--128\relax
\mciteBstWouldAddEndPuncttrue
\mciteSetBstMidEndSepPunct{\mcitedefaultmidpunct}
{\mcitedefaultendpunct}{\mcitedefaultseppunct}\relax
\EndOfBibitem
\bibitem[Cerezo \latin{et~al.}(2021)Cerezo, Arrasmith, Babbush, Benjamin, Endo, Fujii, McClean, Mitarai, Yuan, Cincio, \latin{et~al.} others]{CerArrBab21}
Cerezo,~M.; Arrasmith,~A.; Babbush,~R.; Benjamin,~S.~C.; Endo,~S.; Fujii,~K.; McClean,~J.~R.; Mitarai,~K.; Yuan,~X.; Cincio,~L.; others Variational quantum algorithms. \emph{Nat.\ Rev.\ Phys.} \textbf{2021}, \emph{3}, 625--644\relax
\mciteBstWouldAddEndPuncttrue
\mciteSetBstMidEndSepPunct{\mcitedefaultmidpunct}
{\mcitedefaultendpunct}{\mcitedefaultseppunct}\relax
\EndOfBibitem
\bibitem[Liu \latin{et~al.}(2021)Liu, Li, and Yang]{LiuLiYan21}
Liu,~J.; Li,~Z.; Yang,~J. An efficient adaptive variational quantum solver of the Schr\"odinger equation based on reduced density matrices. \emph{J.~Chem.\ Phys.} \textbf{2021}, \emph{154}, 244112\relax
\mciteBstWouldAddEndPuncttrue
\mciteSetBstMidEndSepPunct{\mcitedefaultmidpunct}
{\mcitedefaultendpunct}{\mcitedefaultseppunct}\relax
\EndOfBibitem
\bibitem[Liu \latin{et~al.}(2022)Liu, Fan, Li, and Yang]{LiuFanLi22}
Liu,~J.; Fan,~Y.; Li,~Z.; Yang,~J. Quantum algorithms for electronic structures: basis sets and boundary conditions. \emph{Chem.\ Soc.\ Rev.} \textbf{2022}, \emph{51}, 3263--3279\relax
\mciteBstWouldAddEndPuncttrue
\mciteSetBstMidEndSepPunct{\mcitedefaultmidpunct}
{\mcitedefaultendpunct}{\mcitedefaultseppunct}\relax
\EndOfBibitem
\bibitem[Huang \latin{et~al.}(2022)Huang, Cai, Li, Ge, Hou, Li, Liu, Shi, Chen, Zheng, \latin{et~al.} others]{HuaCaiLi22}
Huang,~K.; Cai,~X.; Li,~H.; Ge,~Z.-Y.; Hou,~R.; Li,~H.; Liu,~T.; Shi,~Y.; Chen,~C.; Zheng,~D.; others Variational quantum computation of molecular linear response properties on a superconducting quantum processor. \emph{J.~Phys.\ Chem.\ Lett.} \textbf{2022}, \emph{13}, 9114--9121\relax
\mciteBstWouldAddEndPuncttrue
\mciteSetBstMidEndSepPunct{\mcitedefaultmidpunct}
{\mcitedefaultendpunct}{\mcitedefaultseppunct}\relax
\EndOfBibitem
\bibitem[Fedorov \latin{et~al.}(2022)Fedorov, Peng, Govind, and Alexeev]{FedPenGov22}
Fedorov,~D.~A.; Peng,~B.; Govind,~N.; Alexeev,~Y. VQE method: a short survey and recent developments. \emph{Mater. Theory} \textbf{2022}, \emph{6}, 2\relax
\mciteBstWouldAddEndPuncttrue
\mciteSetBstMidEndSepPunct{\mcitedefaultmidpunct}
{\mcitedefaultendpunct}{\mcitedefaultseppunct}\relax
\EndOfBibitem
\bibitem[Bauer \latin{et~al.}(2020)Bauer, Bravyi, Motta, and Chan]{BauBelBra20}
Bauer,~B.; Bravyi,~S.; Motta,~M.; Chan,~G. K.-L. Quantum algorithms for quantum chemistry and quantum materials science. \emph{Chem.\ Rev.} \textbf{2020}, \emph{120}, 12685--12717\relax
\mciteBstWouldAddEndPuncttrue
\mciteSetBstMidEndSepPunct{\mcitedefaultmidpunct}
{\mcitedefaultendpunct}{\mcitedefaultseppunct}\relax
\EndOfBibitem
\bibitem[Kandala \latin{et~al.}(2017)Kandala, Mezzacapo, Temme, Takita, Brink, Chow, and Gambetta]{KanMezTem17}
Kandala,~A.; Mezzacapo,~A.; Temme,~K.; Takita,~M.; Brink,~M.; Chow,~J.~M.; Gambetta,~J.~M. Hardware-efficient variational quantum eigensolver for small molecules and quantum magnets. \emph{Nature} \textbf{2017}, \emph{549}, 242--246\relax
\mciteBstWouldAddEndPuncttrue
\mciteSetBstMidEndSepPunct{\mcitedefaultmidpunct}
{\mcitedefaultendpunct}{\mcitedefaultseppunct}\relax
\EndOfBibitem
\bibitem[Lee \latin{et~al.}(2018)Lee, Huggins, Head-Gordon, and Whaley]{LeeHugHea18}
Lee,~J.; Huggins,~W.~J.; Head-Gordon,~M.; Whaley,~K.~B. Generalized unitary coupled cluster wave functions for quantum computation. \emph{J.~Chem.\ Theory Comput.} \textbf{2018}, \emph{15}, 311--324\relax
\mciteBstWouldAddEndPuncttrue
\mciteSetBstMidEndSepPunct{\mcitedefaultmidpunct}
{\mcitedefaultendpunct}{\mcitedefaultseppunct}\relax
\EndOfBibitem
\bibitem[Ryabinkin \latin{et~al.}(2018)Ryabinkin, Yen, Genin, and Izmaylov]{RyaYenGen18}
Ryabinkin,~I.~G.; Yen,~T.-C.; Genin,~S.~N.; Izmaylov,~A.~F. Qubit coupled cluster method: a systematic approach to quantum chemistry on a quantum computer. \emph{J.~Chem.\ Theory Comput.} \textbf{2018}, \emph{14}, 6317--6326\relax
\mciteBstWouldAddEndPuncttrue
\mciteSetBstMidEndSepPunct{\mcitedefaultmidpunct}
{\mcitedefaultendpunct}{\mcitedefaultseppunct}\relax
\EndOfBibitem
\bibitem[Ryabinkin \latin{et~al.}(2020)Ryabinkin, Lang, Genin, and Izmaylov]{RyaLanCen20}
Ryabinkin,~I.~G.; Lang,~R.~A.; Genin,~S.~N.; Izmaylov,~A.~F. Iterative qubit coupled cluster approach with efficient screening of generators. \emph{J.~Chem.\ Theory Comput.} \textbf{2020}, \emph{16}, 1055--1063\relax
\mciteBstWouldAddEndPuncttrue
\mciteSetBstMidEndSepPunct{\mcitedefaultmidpunct}
{\mcitedefaultendpunct}{\mcitedefaultseppunct}\relax
\EndOfBibitem
\bibitem[Xia and Kais(2020)Xia, and Kais]{XiaKai20}
Xia,~R.; Kais,~S. Qubit coupled cluster singles and doubles variational quantum eigensolver ansatz for electronic structure calculations. \emph{Quant.\ Sci.\ Technol.} \textbf{2020}, \emph{6}, 015001\relax
\mciteBstWouldAddEndPuncttrue
\mciteSetBstMidEndSepPunct{\mcitedefaultmidpunct}
{\mcitedefaultendpunct}{\mcitedefaultseppunct}\relax
\EndOfBibitem
\bibitem[Anand \latin{et~al.}(2022)Anand, Schleich, Alperin-Lea, Jensen, Sim, D{\'\i}az-Tinoco, Kottmann, Degroote, Izmaylov, and Aspuru-Guzik]{AnaSchAlp22}
Anand,~A.; Schleich,~P.; Alperin-Lea,~S.; Jensen,~P.~W.; Sim,~S.; D{\'\i}az-Tinoco,~M.; Kottmann,~J.~S.; Degroote,~M.; Izmaylov,~A.~F.; Aspuru-Guzik,~A. A quantum computing view on unitary coupled cluster theory. \emph{Chem.\ Soc.\ Rev.} \textbf{2022}, \emph{51}, 1659--1684\relax
\mciteBstWouldAddEndPuncttrue
\mciteSetBstMidEndSepPunct{\mcitedefaultmidpunct}
{\mcitedefaultendpunct}{\mcitedefaultseppunct}\relax
\EndOfBibitem
\bibitem[Gard \latin{et~al.}(2020)Gard, Zhu, Barron, Mayhall, Economou, and Barnes]{GarZhuBar20}
Gard,~B.~T.; Zhu,~L.; Barron,~G.~S.; Mayhall,~N.~J.; Economou,~S.~E.; Barnes,~E. Efficient symmetry-preserving state preparation circuits for the variational quantum eigensolver algorithm. \emph{Npj\ Quantum\ Inf.} \textbf{2020}, \emph{6}, 10\relax
\mciteBstWouldAddEndPuncttrue
\mciteSetBstMidEndSepPunct{\mcitedefaultmidpunct}
{\mcitedefaultendpunct}{\mcitedefaultseppunct}\relax
\EndOfBibitem
\bibitem[Fan \latin{et~al.}(2023)Fan, Liu, Li, and Yang]{FanLiuLi23}
Fan,~Y.; Liu,~J.; Li,~Z.; Yang,~J. Quantum circuit matrix product state ansatz for large-scale simulations of molecules. \emph{J.~Chem.\ Theory Comput.} \textbf{2023}, \emph{19}, 5407--5417\relax
\mciteBstWouldAddEndPuncttrue
\mciteSetBstMidEndSepPunct{\mcitedefaultmidpunct}
{\mcitedefaultendpunct}{\mcitedefaultseppunct}\relax
\EndOfBibitem
\bibitem[Zeng \latin{et~al.}(2023)Zeng, Fan, Liu, Li, and Yang]{ZenFanLiu23}
Zeng,~X.; Fan,~Y.; Liu,~J.; Li,~Z.; Yang,~J. Quantum neural network inspired hardware adaptable ansatz for efficient quantum simulation of chemical systems. \emph{J.~Chem.\ Theory Comput.} \textbf{2023}, \emph{19}, 8587--8597\relax
\mciteBstWouldAddEndPuncttrue
\mciteSetBstMidEndSepPunct{\mcitedefaultmidpunct}
{\mcitedefaultendpunct}{\mcitedefaultseppunct}\relax
\EndOfBibitem
\bibitem[Grimsley \latin{et~al.}(2019)Grimsley, Claudino, Economou, Barnes, and Mayhall]{GriClaEco19}
Grimsley,~H.~R.; Claudino,~D.; Economou,~S.~E.; Barnes,~E.; Mayhall,~N.~J. Is the trotterized uccsd ansatz chemically well-defined? \emph{J.~Chem.\ Theory Comput.} \textbf{2019}, \emph{16}, 1--6\relax
\mciteBstWouldAddEndPuncttrue
\mciteSetBstMidEndSepPunct{\mcitedefaultmidpunct}
{\mcitedefaultendpunct}{\mcitedefaultseppunct}\relax
\EndOfBibitem
\bibitem[Romero \latin{et~al.}(2018)Romero, Babbush, McClean, Hempel, Love, and Aspuru-Guzik]{RomBabMcC18}
Romero,~J.; Babbush,~R.; McClean,~J.~R.; Hempel,~C.; Love,~P.~J.; Aspuru-Guzik,~A. Strategies for quantum computing molecular energies using the unitary coupled cluster ansatz. \emph{Quant.\ Sci.\ Technol.} \textbf{2018}, \emph{4}, 014008\relax
\mciteBstWouldAddEndPuncttrue
\mciteSetBstMidEndSepPunct{\mcitedefaultmidpunct}
{\mcitedefaultendpunct}{\mcitedefaultseppunct}\relax
\EndOfBibitem
\bibitem[Barkoutsos \latin{et~al.}(2018)Barkoutsos, Gonthier, Sokolov, Moll, Salis, Fuhrer, Ganzhorn, Egger, Troyer, Mezzacapo, \latin{et~al.} others]{BarGonSok18}
Barkoutsos,~P.~K.; Gonthier,~J.~F.; Sokolov,~I.; Moll,~N.; Salis,~G.; Fuhrer,~A.; Ganzhorn,~M.; Egger,~D.~J.; Troyer,~M.; Mezzacapo,~A.; others Quantum algorithms for electronic structure calculations: Particle-hole Hamiltonian and optimized wave-function expansions. \emph{Phys.\ Rev.~A} \textbf{2018}, \emph{98}, 022322\relax
\mciteBstWouldAddEndPuncttrue
\mciteSetBstMidEndSepPunct{\mcitedefaultmidpunct}
{\mcitedefaultendpunct}{\mcitedefaultseppunct}\relax
\EndOfBibitem
\bibitem[Grimsley \latin{et~al.}(2019)Grimsley, Economou, Barnes, and Mayhall]{GriEcoBar19}
Grimsley,~H.~R.; Economou,~S.~E.; Barnes,~E.; Mayhall,~N.~J. An adaptive variational algorithm for exact molecular simulations on a quantum computer. \emph{Nat.\ Chem.} \textbf{2019}, \emph{10}, 3007\relax
\mciteBstWouldAddEndPuncttrue
\mciteSetBstMidEndSepPunct{\mcitedefaultmidpunct}
{\mcitedefaultendpunct}{\mcitedefaultseppunct}\relax
\EndOfBibitem
\bibitem[Tang \latin{et~al.}(2021)Tang, Shkolnikov, Barron, Grimsley, Mayhall, Barnes, and Economou]{TanShkBar21}
Tang,~H.~L.; Shkolnikov,~V.; Barron,~G.~S.; Grimsley,~H.~R.; Mayhall,~N.~J.; Barnes,~E.; Economou,~S.~E. Qubit-ADAPT-VQE: An Adaptive Algorithm for Constructing Hardware-Efficient Ans\"atze on a Quantum Processor. \emph{PRX Quantum} \textbf{2021}, \emph{2}, 020310\relax
\mciteBstWouldAddEndPuncttrue
\mciteSetBstMidEndSepPunct{\mcitedefaultmidpunct}
{\mcitedefaultendpunct}{\mcitedefaultseppunct}\relax
\EndOfBibitem
\bibitem[Yordanov \latin{et~al.}(2020)Yordanov, Arvidsson-Shukur, and Barnes]{YorArvBar20}
Yordanov,~Y.~S.; Arvidsson-Shukur,~D. R.~M.; Barnes,~C. H.~W. Efficient quantum circuits for quantum computational chemistry. \emph{Phys.\ Rev.~A} \textbf{2020}, \emph{102}, 062612\relax
\mciteBstWouldAddEndPuncttrue
\mciteSetBstMidEndSepPunct{\mcitedefaultmidpunct}
{\mcitedefaultendpunct}{\mcitedefaultseppunct}\relax
\EndOfBibitem
\bibitem[Yordanov \latin{et~al.}(2021)Yordanov, Armaos, Barnes, and Arvidsson-Shukur]{YorYorArm21}
Yordanov,~Y.~S.; Armaos,~V.; Barnes,~C.~H.; Arvidsson-Shukur,~D.~R. Qubit-excitation-based adaptive variational quantum eigensolver. \emph{Commun. Phys.} \textbf{2021}, \emph{4}, 228\relax
\mciteBstWouldAddEndPuncttrue
\mciteSetBstMidEndSepPunct{\mcitedefaultmidpunct}
{\mcitedefaultendpunct}{\mcitedefaultseppunct}\relax
\EndOfBibitem
\bibitem[Magoulas and Evangelista(2023)Magoulas, and Evangelista]{MagEva23}
Magoulas,~I.; Evangelista,~F.~A. CNOT-Efficient Circuits for Arbitrary Rank Many-Body Fermionic and Qubit Excitations. \emph{J.~Chem.\ Theory Comput.} \textbf{2023}, \emph{19}, 822--836\relax
\mciteBstWouldAddEndPuncttrue
\mciteSetBstMidEndSepPunct{\mcitedefaultmidpunct}
{\mcitedefaultendpunct}{\mcitedefaultseppunct}\relax
\EndOfBibitem
\bibitem[Magoulas and Evangelista(2023)Magoulas, and Evangelista]{MagEva23b}
Magoulas,~I.; Evangelista,~F.~A. Linear-Scaling Quantum Circuits for Computational Chemistry. \emph{J.~Chem.\ Theory Comput.} \textbf{2023}, \emph{19}, 4815--4821\relax
\mciteBstWouldAddEndPuncttrue
\mciteSetBstMidEndSepPunct{\mcitedefaultmidpunct}
{\mcitedefaultendpunct}{\mcitedefaultseppunct}\relax
\EndOfBibitem
\bibitem[Jordan and Wigner(1928)Jordan, and Wigner]{JorWig28}
Jordan,~P.; Wigner,~E. {\"U}ber das Paulische {\"A}quivalenzverbot. \emph{Eur. Phys. J. A} \textbf{1928}, \emph{47}, 631--651\relax
\mciteBstWouldAddEndPuncttrue
\mciteSetBstMidEndSepPunct{\mcitedefaultmidpunct}
{\mcitedefaultendpunct}{\mcitedefaultseppunct}\relax
\EndOfBibitem
\bibitem[Bravyi and Kitaev(2002)Bravyi, and Kitaev]{BraKit02}
Bravyi,~S.; Kitaev,~A. Fermionic quantum computation. \emph{Ann. Phys.} \textbf{2002}, \emph{298}, 210--226\relax
\mciteBstWouldAddEndPuncttrue
\mciteSetBstMidEndSepPunct{\mcitedefaultmidpunct}
{\mcitedefaultendpunct}{\mcitedefaultseppunct}\relax
\EndOfBibitem
\bibitem[Seeley \latin{et~al.}(2012)Seeley, Richard, and Love]{SeeRicLov12}
Seeley,~J.~T.; Richard,~M.~J.; Love,~P.~J. The Bravyi-Kitaev transformation for quantum computation of electronic structure. \emph{J.~Chem.\ Phys.} \textbf{2012}, \emph{137}\relax
\mciteBstWouldAddEndPuncttrue
\mciteSetBstMidEndSepPunct{\mcitedefaultmidpunct}
{\mcitedefaultendpunct}{\mcitedefaultseppunct}\relax
\EndOfBibitem
\bibitem[McClean \latin{et~al.}(2017)McClean, Kimchi-Schwartz, Carter, and De~Jong]{McCKimCar17}
McClean,~J.~R.; Kimchi-Schwartz,~M.~E.; Carter,~J.; De~Jong,~W.~A. Hybrid quantum-classical hierarchy for mitigation of decoherence and determination of excited states. \emph{Phys.\ Rev.~A} \textbf{2017}, \emph{95}, 042308\relax
\mciteBstWouldAddEndPuncttrue
\mciteSetBstMidEndSepPunct{\mcitedefaultmidpunct}
{\mcitedefaultendpunct}{\mcitedefaultseppunct}\relax
\EndOfBibitem
\bibitem[McClean \latin{et~al.}(2020)McClean, Jiang, Rubin, Babbush, and Neven]{McCJiaRub20}
McClean,~J.~R.; Jiang,~Z.; Rubin,~N.~C.; Babbush,~R.; Neven,~H. Decoding quantum errors with subspace expansions. \emph{Nat.\ Chem.} \textbf{2020}, \emph{11}, 636\relax
\mciteBstWouldAddEndPuncttrue
\mciteSetBstMidEndSepPunct{\mcitedefaultmidpunct}
{\mcitedefaultendpunct}{\mcitedefaultseppunct}\relax
\EndOfBibitem
\bibitem[Ollitrault \latin{et~al.}(2020)Ollitrault, Kandala, Chen, Barkoutsos, Mezzacapo, Pistoia, Sheldon, Woerner, Gambetta, and Tavernelli]{OllKanChe20}
Ollitrault,~P.~J.; Kandala,~A.; Chen,~C.-F.; Barkoutsos,~P.~K.; Mezzacapo,~A.; Pistoia,~M.; Sheldon,~S.; Woerner,~S.; Gambetta,~J.~M.; Tavernelli,~I. Quantum equation of motion for computing molecular excitation energies on a noisy quantum processor. \emph{Phys.\ Rev.\ Research} \textbf{2020}, \emph{2}, 043140\relax
\mciteBstWouldAddEndPuncttrue
\mciteSetBstMidEndSepPunct{\mcitedefaultmidpunct}
{\mcitedefaultendpunct}{\mcitedefaultseppunct}\relax
\EndOfBibitem
\bibitem[Fan \latin{et~al.}(2021)Fan, Liu, Li, and Yang]{FanLiuLi21}
Fan,~Y.; Liu,~J.; Li,~Z.; Yang,~J. Equation-of-motion theory to calculate accurate band structures with a quantum computer. \emph{J.~Phys.\ Chem.\ Lett.} \textbf{2021}, \emph{12}, 8833--8840\relax
\mciteBstWouldAddEndPuncttrue
\mciteSetBstMidEndSepPunct{\mcitedefaultmidpunct}
{\mcitedefaultendpunct}{\mcitedefaultseppunct}\relax
\EndOfBibitem
\bibitem[Asthana \latin{et~al.}(2023)Asthana, Kumar, Abraham, Grimsley, Zhang, Cincio, Tretiak, Dub, Economou, Barnes, \latin{et~al.} others]{AstKumAbr23}
Asthana,~A.; Kumar,~A.; Abraham,~V.; Grimsley,~H.; Zhang,~Y.; Cincio,~L.; Tretiak,~S.; Dub,~P.~A.; Economou,~S.~E.; Barnes,~E.; others Quantum self-consistent equation-of-motion method for computing molecular excitation energies, ionization potentials, and electron affinities on a quantum computer. \emph{Chem. Sci.} \textbf{2023}, \emph{14}, 2405--2418\relax
\mciteBstWouldAddEndPuncttrue
\mciteSetBstMidEndSepPunct{\mcitedefaultmidpunct}
{\mcitedefaultendpunct}{\mcitedefaultseppunct}\relax
\EndOfBibitem
\bibitem[Nakanishi \latin{et~al.}(2019)Nakanishi, Mitarai, and Fujii]{NakMitFuj19}
Nakanishi,~K.~M.; Mitarai,~K.; Fujii,~K. Subspace-search variational quantum eigensolver for excited states. \emph{Phys.\ Rev.\ Research} \textbf{2019}, \emph{1}, 033062\relax
\mciteBstWouldAddEndPuncttrue
\mciteSetBstMidEndSepPunct{\mcitedefaultmidpunct}
{\mcitedefaultendpunct}{\mcitedefaultseppunct}\relax
\EndOfBibitem
\bibitem[Higgott \latin{et~al.}(2019)Higgott, Wang, and Brierley]{HigOscWan19}
Higgott,~O.; Wang,~D.; Brierley,~S. Variational quantum computation of excited states. \emph{Quantum} \textbf{2019}, \emph{3}, 156\relax
\mciteBstWouldAddEndPuncttrue
\mciteSetBstMidEndSepPunct{\mcitedefaultmidpunct}
{\mcitedefaultendpunct}{\mcitedefaultseppunct}\relax
\EndOfBibitem
\bibitem[Yordanov \latin{et~al.}(2022)Yordanov, Barnes, and Arvidsson-Shukur]{YorBarArv22}
Yordanov,~Y.~S.; Barnes,~C. H.~W.; Arvidsson-Shukur,~D. R.~M. Molecular-excited-state calculations with the qubit-excitation-based adaptive variational quantum eigensolver protocol. \emph{Phys.\ Rev.~A} \textbf{2022}, \emph{106}, 032434\relax
\mciteBstWouldAddEndPuncttrue
\mciteSetBstMidEndSepPunct{\mcitedefaultmidpunct}
{\mcitedefaultendpunct}{\mcitedefaultseppunct}\relax
\EndOfBibitem
\bibitem[Chen \latin{et~al.}(2021)Chen, Cheng, and Freericks]{CheCheFre21}
Chen,~J.; Cheng,~H.-P.; Freericks,~J.~K. Quantum-inspired algorithm for the factorized form of unitary coupled cluster theory. \emph{J.~Chem.\ Theory Comput.} \textbf{2021}, \emph{17}, 841--847\relax
\mciteBstWouldAddEndPuncttrue
\mciteSetBstMidEndSepPunct{\mcitedefaultmidpunct}
{\mcitedefaultendpunct}{\mcitedefaultseppunct}\relax
\EndOfBibitem
\bibitem[Fan \latin{et~al.}(2022)Fan, Liu, Zeng, Xu, Shang, Li, and Yang]{FanLiuZen22}
Fan,~Y.; Liu,~J.; Zeng,~X.; Xu,~Z.; Shang,~H.; Li,~Z.; Yang,~J. Q$^2$Chemistry: A quantum computation platform for quantum chemistry. \emph{arXiv} \textbf{2022}, 2208.10978\relax
\mciteBstWouldAddEndPuncttrue
\mciteSetBstMidEndSepPunct{\mcitedefaultmidpunct}
{\mcitedefaultendpunct}{\mcitedefaultseppunct}\relax
\EndOfBibitem
\bibitem[Sun \latin{et~al.}(2018)Sun, Berkelbach, Blunt, Booth, Guo, Li, Liu, McClain, Sayfutyarova, Sharma, Wouters, and Chan]{SunBerBlu18}
Sun,~Q.; Berkelbach,~T.~C.; Blunt,~N.~S.; Booth,~G.~H.; Guo,~S.; Li,~Z.; Liu,~J.; McClain,~J.~D.; Sayfutyarova,~E.~R.; Sharma,~S.; Wouters,~S.; Chan,~G. K.-L. PYSCF: the Python-based simulations of chemistry framework. \emph{Wiley Interdiscip. Rev.: Comput. Mol. Sci.} \textbf{2018}, \emph{8}\relax
\mciteBstWouldAddEndPuncttrue
\mciteSetBstMidEndSepPunct{\mcitedefaultmidpunct}
{\mcitedefaultendpunct}{\mcitedefaultseppunct}\relax
\EndOfBibitem
\bibitem[McClean \latin{et~al.}(2020)McClean, Rubin, Sung, Kivlichan, Bonet-Monroie, Cao, Dai, Fried, Gidney, Gimby, Gokhale, Haner, Hardikar, Havlicek, Higgott, Huang, Izaac, Jiang, Liu, McArdle, Neeley, O'Brien, O'Gorman, Ozfidan, Radin, Romero, Sawaya, Senjean, Setia, Sim, Steiger, Steudtner, Sun, Sun, Wang, Zhang, and Babbush]{McCRubSun20}
McClean,~J.~R.; Rubin,~N.~C.; Sung,~K.~J.; Kivlichan,~I.~D.; Bonet-Monroie,~X.; Cao,~Y.; Dai,~C.; Fried,~E.~S.; Gidney,~C.; Gimby,~B.; Gokhale,~P.; Haner,~T.; Hardikar,~T.; Havlicek,~V.; Higgott,~O.; Huang,~C.; Izaac,~J.; Jiang,~Z.; Liu,~X.; McArdle,~S.; Neeley,~M.; O'Brien,~T.; O'Gorman,~B.; Ozfidan,~I.; Radin,~M.~D.; Romero,~J.; Sawaya,~N. P.~D.; Senjean,~B.; Setia,~K.; Sim,~S.; Steiger,~D.~S.; Steudtner,~M.; Sun,~Q.; Sun,~W.; Wang,~D.; Zhang,~F.; Babbush,~R. OpenFermion: the electronic structure package for quantum computers. \emph{Quant.\ Sci.\ Technol.} \textbf{2020}, \emph{5}\relax
\mciteBstWouldAddEndPuncttrue
\mciteSetBstMidEndSepPunct{\mcitedefaultmidpunct}
{\mcitedefaultendpunct}{\mcitedefaultseppunct}\relax
\EndOfBibitem
\bibitem[Virtanen \latin{et~al.}(2020)Virtanen, Gommers, Oliphant, Haberland, Reddy, Cournapeau, Burovski, Peterson, Weckesser, Bright, van~der Walt, Brett, Wilson, Millman, Mayorov, Nelson, Jones, Kern, Larson, Carey, Polat, Feng, Moore, VanderPlas, Laxalde, Perktold, Cimrman, Henriksen, Quintero, Harris, Archibald, Ribeiro, Pedregosa, van Mulbregt, and Contributors]{VirGomOli20}
Virtanen,~P.; Gommers,~R.; Oliphant,~T.~E.; Haberland,~M.; Reddy,~T.; Cournapeau,~D.; Burovski,~E.; Peterson,~P.; Weckesser,~W.; Bright,~J.; van~der Walt,~S.~J.; Brett,~M.; Wilson,~J.; Millman,~K.~J.; Mayorov,~N.; Nelson,~A. R.~J.; Jones,~E.; Kern,~R.; Larson,~E.; Carey,~C.~J.; Polat,~I.; Feng,~Y.; Moore,~E.~W.; VanderPlas,~J.; Laxalde,~D.; Perktold,~J.; Cimrman,~R.; Henriksen,~I.; Quintero,~E.~A.; Harris,~C.~R.; Archibald,~A.~M.; Ribeiro,~A.~H.; Pedregosa,~F.; van Mulbregt,~P.; Contributors,~S. .~. SciPy 1.0: fundamental algorithms for scientific computing in Python. \emph{Nat. Methods} \textbf{2020}, \emph{17}, 261--272\relax
\mciteBstWouldAddEndPuncttrue
\mciteSetBstMidEndSepPunct{\mcitedefaultmidpunct}
{\mcitedefaultendpunct}{\mcitedefaultseppunct}\relax
\EndOfBibitem
\bibitem[Liu \latin{et~al.}(2020)Liu, Wan, Li, and Yang]{LiuWanLi20}
Liu,~J.; Wan,~L.; Li,~Z.; Yang,~J. Simulating Periodic Systems on a Quantum Computer Using Molecular Orbitals. \emph{J.~Chem.\ Theory Comput.} \textbf{2020}, \emph{16}, 6904--6914\relax
\mciteBstWouldAddEndPuncttrue
\mciteSetBstMidEndSepPunct{\mcitedefaultmidpunct}
{\mcitedefaultendpunct}{\mcitedefaultseppunct}\relax
\EndOfBibitem
\end{mcitethebibliography}
}

\vspace{3ex}

\section*{Appendix A. Circuit decomposition}
        
    The two-body sQEB excitation operator is $U^{pq}_{rs}(\theta)=\exp{(\theta \tau^{pq}_{rs})}$ and the corresponding generator is
    \begin{equation}
    \begin{split}
        \tau^{pq}_{rs}
        &= - \frac i4(X_pY_qX_rX_s+Y_pY_qY_rX_s-X_pX_qX_rY_s-Y_pX_qY_rY_s).  \\
        &= (Q^{\dag}_p Q^{\dag}_q Q_r Q_s + Q^{\dag}_q Q^{\dag}_r Q_s Q_p) - h.c.
    \end{split}
    \end{equation}
    A multi-qubit controlled rotation gate $R_y(\theta,\left\{ q_1...q_m\right\},q_0)$, in conjunction with several CNOT gates, can be utilized to implement this unitary operation.
    
    \subsection*{A1. Multi-qubit-controlled rotation gate}

        The Pauli-Y operator is $\begin{pmatrix}0 & -i \\ i & 0 \end{pmatrix}$, and $R_y$ gate is $R_y(\theta) = e^{-i\theta Y} = \begin{pmatrix} \cos{\theta} & -\sin{\theta} \\ \sin{\theta} & \cos{\theta} \end{pmatrix}$, the corresponding generator is $-iY = \begin{pmatrix}0 & -1 \\ 1 & 0 \end{pmatrix}$
    
        The matrix of a multi-qubit-controlled rotation gate $R_y(\theta,\left\{ q_0,q_1\right\},q_2)$ is shown below (the basis is of $\ket{q_0q_1q_2}$ convention)
        \begin{equation}\label{Eq:matrix1}
        \bordermatrix{%
                   & \ket{000}  & \ket{001}  &\ket{010}  & \ket{011}  & \ket{100}  &\ket{101}  & \ket{110}  & \ket{111}\cr
        \ket{000}  &1  &0  &0  &0  &0  &0  &0  &0 \cr
        \ket{001}  &0  &1  &0  &0  &0  &0  &0  &0 \cr
        \ket{010}  &0  &0  &1  &0  &0  &0  &0  &0 \cr
        \ket{011}  &0  &0  &0  &1  &0  &0  &0  &0 \cr
        \ket{100}  &0  &0  &0  &0  &1  &0  &0  &0 \cr
        \ket{101}  &0  &0  &0  &0  &0  &1  &0  &0 \cr
        \ket{110}  &0  &0  &0  &0  &0  &0  &\cos{\theta}  &-\sin{\theta} \cr
        \ket{111}  &0  &0  &0  &0  &0  &0  &\sin{\theta}  &\cos{\theta}
        }
        \end{equation}
        This unitary only mix two basis $\ket{110}$ and $\ket{111}$ that satisfy $q_0=q_1=1$.  We can show it in a simplified form by ignoring the ones in diagonal and all zeros,
        \begin{equation}\label{Eq:matrix1}
        \bordermatrix{%
                   & \ket{110}    & \ket{111}\cr
        \ket{110}  &\cos{\theta}  &-\sin{\theta} \cr
        \ket{111}  &\sin{\theta}  &\cos{\theta}
        }
        \end{equation}
        The generator of this unitary is $-iP_0^{(1)}P_1^{(1)}Y_2$ ($P_i^{(1)}$ is the projector $\ket{1}\bra{1}$ in i-th qubit), and the matrix (ignoring zeros) is 

        \begin{equation}\label{Eq:matrix1}
        \bordermatrix{%
                   & \ket{110}    & \ket{111}\cr
        \ket{110}  &0  &-1 \cr
        \ket{111}  &1  &0
        }
        \end{equation}

    \subsection*{A2. Circuit decomposition}
    
        Returning to our generator $\tau^{pq}_{rs}$, its matrix representation is provided below, with the basis in the $\ket{q_pq_qq_rq_s}$ convention.
        \begin{equation}\label{Eq:matrix1}
        \bordermatrix{%
                   & \ket{0011}    & \ket{1100}  & \ket{1001}  & \ket{0110}\cr
        \ket{0011}  &0  &-1 &0  &0  \cr
        \ket{1100}  &1  &0  &0  &0  \cr
        \ket{1001}  &0  &0  &0  &-1 \cr
        \ket{0110}  &0  &0  &1  &0
        }
        \end{equation}

        We can see that it mixes two pairs of bases, each containing two particles, thereby preserving the particle number symmetry. By imposing the conditions $(\sigma_p+\sigma_q=\sigma_r+\sigma_s) \wedge (\sigma_q+\sigma_r=\sigma_p+\sigma_s)$, the excitation will preserve the $S_z$ symmetry. This condition simplifies to $(\sigma_p=\sigma_r) \wedge (\sigma_q=\sigma_s)$, where $\sigma_i$ represents the spin of the i-th spin-orbital.

        We can chooce a $R_y(\theta,\left\{ q_r,q_s\right\},q_q)$ gate as the building block. The generator of this gate in the four qubit space $\left\{ \ket{q_pq_qq_rq_s}\right\}$ is given by

        \begin{equation}\label{Eq:matrix1}
        \bordermatrix{%
                   & \ket{0011}    & \ket{0111}  & \ket{1011}  & \ket{1111}\cr
        \ket{0011}  &0  &-1 &0  &0  \cr
        \ket{0111}  &1  &0  &0  &0  \cr
        \ket{1011}  &0  &0  &0  &-1 \cr
        \ket{1111}  &0  &0  &1  &0
        }
        \end{equation}

        The two generators have a similar form, differing only in the exchange between different bases. To obtain the $\tau^{pq}_{rs}$ generator and the corresponding unitary operation, we can apply some CNOT gates to transform the basis. The resulting circuit is shown in the figure below.

            \begin{center}
            \noindent\makebox[\textwidth]{
            \begin{quantikz}
                &\lstick{$p$} &\ctrl{2} &\qw       &\targ{}   &\qw   &\targ{}   &\qw       &\ctrl{2} &\qw\\
                &\lstick{$q$} &\qw       &\ctrl{2} &\ctrl{-1} & \gate{R_y(\theta)} &\ctrl{-1} &\ctrl{2} &\qw       &\qw\\
                &\lstick{$r$} &\targ{}   &\qw       &\qw       &\ctrl{-1}            &\qw       &\qw       &\targ{}   &\qw\\
                &\lstick{$s$} &\qw       &\targ{}   &\qw       &\ctrl{-1}            &\qw       &\targ{}   &\qw       &\qw
            \end{quantikz}
            }\end{center}

        By utilizing the construction circuit for $R_y(\theta,\left\{ q_0,q_1\right\},q_2)$ along with the circuit identity (ignoring global phase), one CNOT gate can be eliminated. Consequently, the final circuit requires 9 CNOT gates and has a total circuit depth of 7.

            \begin{center}
            \noindent\makebox[\textwidth]{
            \begin{quantikz}
                &\ctrl{2} &\qw \\
                &\ctrl{1} &\qw \\
                & \gate{R_y(\theta)} &\qw
            \end{quantikz} = 
            \begin{quantikz}
                &\qw &\qw &\qw &\ctrl{2} &\qw &\qw &\qw &\ctrl{2} &\qw \\
                &\qw &\ctrl{1} &\qw &\qw &\qw &\ctrl{1} &\qw &\qw &\qw \\
                &\gate{R_y(\frac{\theta}{4})} &\control{} &\gate{R_y(-\frac{\theta}{4})} &\control{} &\gate{R_y(\frac{\theta}{4})} &\control{} &\gate{R_y(-\frac{\theta}{4})} &\control{} &\qw
            \end{quantikz}
            }\end{center}

            \begin{center}
            \noindent\makebox[\textwidth]{
            \begin{quantikz}
                &\ctrl{1}   &\targ{} &\qw \\
                &\control{} &\ctrl{-1}  &\qw
            \end{quantikz}
            = \begin{quantikz}
                &\gate{R_y(\frac{\pi}{2})} &\gate{R_z(\frac{\pi}{2})} &\targ{} &\gate{R_z(-\frac{\pi}{2})} &\gate{R_y(-\frac{\pi}{2})} &\qw \\
                &\qw &\qw &\ctrl{-1} &\gate{R_z(\frac{\pi}{2})}  &\qw &\qw
            \end{quantikz}
            }\end{center}

        The result circuit is

            \begin{center}
            \noindent\makebox[\textwidth]{
            \begin{quantikz}
                &\lstick{$p$} &\ctrl{2} &\qw       &\targ{} 
                    &\qw &\qw &\qw &\qw &\qw &\qw &\qw &\qw 
                        &\targ{} &\qw &\ctrl{2} &\qw\\
                &\lstick{$q$} &\qw       &\ctrl{2} &\ctrl{-1} 
                    &\gate{R_y(\frac{\theta}{4})} &\control{} &\gate{R_y(-\frac{\theta}{4})} &\control{} &\gate{R_y(\frac{\theta}{4})} &\control{} &\gate{R_y(-\frac{\theta}{4})} &\control{} 
                        &\ctrl{-1} &\ctrl{2} &\qw &\qw\\
                &\lstick{$r$} &\targ{}   &\qw       &\qw 
                    &\qw &\ctrl{-1} &\qw &\qw &\qw &\ctrl{-1} &\qw &\qw   &\qw &\qw &\targ{} &\qw\\
                &\lstick{$s$} &\qw       &\targ{}   &\qw 
                    &\qw &\qw &\qw &\ctrl{-2} &\qw &\qw &\qw &\ctrl{-2}  &\qw &\targ{} &\qw &\qw
            \end{quantikz}
            }\end{center}

        Then apply the circuit identity

            \begin{center}
            \noindent\makebox[\textwidth]{
            \begin{quantikz}[column sep = 0.3cm]
                &\lstick{$p$} &\ctrl{2} &\qw       &\targ{} 
                    &\qw &\qw &\qw &\qw &\qw &\qw &\qw &\targ{} 
                        &\qw  &\ctrl{2} &\qw &\qw \\
                &\lstick{$q$} &\qw       &\ctrl{2} &\ctrl{-1} 
                    &\gate{R_y(\frac{\theta}{4})} &\control{} &\gate{R_y(-\frac{\theta}{4})} &\control{} &\gate{R_y(\frac{\theta}{4})} &\control{} &\gate{R_y(-\frac{\theta}{4})} &\ctrl{-1} 
                        &\ctrl{2} &\qw &\gate{R_z(\frac{\pi}{2})} &\qw  \\
                &\lstick{$r$} &\targ{}   &\qw       &\qw 
                    &\qw &\ctrl{-1} &\qw &\qw &\qw &\ctrl{-1} &\qw &\qw &\qw  &\targ{} &\qw &\qw \\
                &\lstick{$s$} &\qw       &\targ{}   &\qw 
                    &\qw &\qw &\qw &\ctrl{-2} 
                        &\gate{R_y(\frac{\pi}{2})} &\qw &\gate{R_z(\frac{\pi}{2})} &\qw &\targ{} &\gate{R_z(-\frac{\pi}{2})} &\gate{R_y(-\frac{\pi}{2})} &\qw
            \end{quantikz}
            }\end{center}

        We will need 9 CNOTs and 7 layer of CNOTs to achieve such operator.

\section*{Appendix B. Other circuits for QEB and sQEB excitation}

    \subsection*{B1. QEB excitation}

    Actually, quantum circuit that realize the QEB double excitation $U(\theta) = e^{\theta \kappa}$ is not unique, where $\kappa^{pq}_{rs} = Q^\dag_p Q^\dag_q Q_r  Q_s - h.c.$. Based on definition of $\kappa^{pq}_{rs}$, we can see 
    \begin{equation}
        \kappa^{pq}_{rs} = \kappa^{qp}_{rs} = \kappa^{pq}_{sr} = \kappa^{qp}_{sr} = 
        -\kappa^{rs}_{pq} = -\kappa^{sr}_{pq} = -\kappa^{rs}_{qp} = -\kappa^{sr}_{qp}
    \end{equation}
    from which we can exchange the indices $p,q,r,s$ to get some equivalent circuit. We do the exchange $p \leftrightarrow q, r \leftrightarrow s$, connect below two circuits.

    \begin{center}
    \noindent\makebox[\textwidth]{
    \begin{quantikz}
        &\lstick{$p$} &\ctrl{1}   &\ctrl{2} & \gate{R_y(\theta)} &\ctrl{2} &\ctrl{1}      &\qw \\
        &\lstick{$q$} &\targ{}     &\qw       &\octrl{-1}              &\qw       &\targ{}        &\qw\\
        &\lstick{$r$} &\ctrl{1}   &\targ{}   &\ctrl{-1}             &\targ{}   &\ctrl{1}      &\qw\\ 
        &\lstick{$s$} &\targ{}     &\qw       &\octrl{-1}           &\qw       &\targ{}        &\qw
    \end{quantikz}
    =\begin{quantikz}
        &\lstick{$p$} &\targ{}     &\qw       &\octrl{1}           &\qw       &\targ{}        &\qw\\
        &\lstick{$q$} &\ctrl{-1}   &\ctrl{2}  &\gate{R_y(\theta)}    &\ctrl{2}  &\ctrl{-1}      &\qw\\ 
        &\lstick{$r$} &\targ{}     &\qw       &\octrl{-1}             &\qw       &\targ{}        &\qw\\
        &\lstick{$s$} &\ctrl{-1}   &\targ{}   &\ctrl{-1}              &\targ{}   &\ctrl{-1}      &\qw
    \end{quantikz}
    }\end{center}

    Apart from these trivial circuits, there are other circuit can implement the QEB double excitation, they all have the similar structure, multi-qubit controlled rotation gate sandwiched by CNOT layers.
    
    \begin{center}
    \noindent\makebox[\textwidth]{
    \begin{quantikz}
        &\lstick{$p$} &\ctrl{2} &\qw        &\targ{}   &\octrl{1} &\targ{}  &\qw       &\ctrl{2} &\qw\\
        &\lstick{$q$} &\qw       &\ctrl{2}  &\ctrl{-1}  &\gate{R_y(\theta)}   &\ctrl{-1}  &\ctrl{2}  &\qw       &\qw \\
        &\lstick{$r$} &\targ{}   &\qw       &\qw       &\ctrl{-1}                 &\qw        &\qw       &\targ{}   &\qw\\
        &\lstick{$s$} &\qw       &\targ{}   &\qw       &\ctrl{-1}                 &\qw        &\targ{}   &\qw       &\qw
    \end{quantikz}
    \begin{quantikz}
        &\lstick{$p$} &\qw       &\ctrl{2}  &\ctrl{3}  &\gate{R_y(\theta)}   &\ctrl{3}  &\ctrl{2}  &\qw       &\qw \\
        &\lstick{$q$} &\targ{} &\qw        &\qw   &\ctrl{-1} &\qw  &\qw       &\targ{} &\qw\\
        &\lstick{$r$} &\qw       &\targ{}   &\qw       &\ctrl{-1}                 &\qw        &\targ{}   &\qw       &\qw\\
        &\lstick{$s$} &\ctrl{-2}   &\qw      &\targ{}       &\ctrl{-1}                 &\targ{}        &\qw       &\ctrl{-2}   &\qw
    \end{quantikz}
    }\end{center}

    \begin{center}
    \noindent\makebox[\textwidth]{
    \begin{quantikz}
        &\lstick{$p$} &\qw       &\qw  &\ctrl{3}  &\gate{R_y(\theta)}   &\ctrl{3}  &\qw  &\qw   &\qw \\
        &\lstick{$q$} &\qw    &\targ{} &\qw  &\ctrl{-1}  &\qw &\targ{}  &\qw    &\qw\\
        &\lstick{$r$} &\targ{}   &\qw   &\qw   &\octrl{-1}                 &\qw   &\qw    &\targ{}   &\qw\\
        &\lstick{$s$} &\ctrl{-1}   &\ctrl{-2}       &\targ{}       &\ctrl{-1}     &\targ{}        &\ctrl{-2}       &\ctrl{-1}   &\qw\\
    \end{quantikz}
    }\end{center}
    
    For one circuit, the choice of CNOT layer is not unique too. An example is provided  below.
    
    \begin{center}
    \noindent\makebox[\textwidth]{
    \begin{quantikz}
        &\lstick{$p$} &\qw   &\qw &\ctrl{1}  &\gate{R_y(\theta)} &\ctrl{1} &\qw  &\qw    &\qw \\
        &\lstick{$q$} &\qw     &\ctrl{1}     &\targ{}  &\octrl{-1}              &\targ{}  &\ctrl{1} &\qw     &\qw\\
        &\lstick{$r$} &\ctrl{1}   &\targ{}   &\qw  &\ctrl{-1}             &\qw  &\targ{}   &\ctrl{1}      &\qw\\ 
        &\lstick{$s$} &\targ{}     &\qw       &\qw  &\octrl{-1}           &\qw &\qw     &\targ{}        &\qw
    \end{quantikz}
    }\end{center}
    
    One can take advantageous of these flexibility when construct the global circuit.

    \subsection*{B2. sQEB excitation}

    We have proposed the quantum circuit for two-body sQEB excitation 
    \begin{center}
    \noindent\makebox[\textwidth]{
    \begin{quantikz}
        &\lstick{$p$} &\ctrl{2} &\qw       &\targ{}   &\qw   &\targ{}   &\qw       &\ctrl{2} &\qw\\
        &\lstick{$q$} &\qw       &\ctrl{2} &\ctrl{-1} & \gate{R_y(\theta)} &\ctrl{-1} &\ctrl{2} &\qw       &\qw\\
        &\lstick{$r$} &\targ{}   &\qw       &\qw       &\ctrl{-1}            &\qw       &\qw       &\targ{}   &\qw\\
        &\lstick{$s$} &\qw       &\targ{}   &\qw       &\ctrl{-1}            &\qw       &\targ{}   &\qw       &\qw
    \end{quantikz}
    }\end{center}
    Quantum circuit that realize the sQEB double excitation $U(\theta) = e^{\theta \tau}$ is not unique too. Here $\tau^{pq}_{rs} = (Q^\dag_p Q^\dag_q Q_r  Q_s + Q^\dag_q Q^\dag_r Q_s Q_p) - h.c.$. Based on definition of $\tau^{pq}_{rs}$, we can see 
    \begin{equation}
        \tau^{pq}_{rs} = \tau^{rq}_{ps} = -\tau^{ps}_{rq} = -\tau^{rs}_{pq}
    \end{equation}
    from which we can exchange the indices $p,q,r,s$ to get some equivalent circuit.

    Apart from these trivially equivalent circuits, there only exist two equivalent circuit that differ by the CNOT layer.
    \begin{center}
    \noindent\makebox[\textwidth]{
    \begin{quantikz}
        &\lstick{$p$} &\ctrl{2} &\targ{}       &\qw   &\qw                 &\qw   &\targ{}       &\ctrl{2} &\qw\\
        &\lstick{$q$} &\qw       &\qw &\ctrl{2} & \gate{R_y(\theta)} &\ctrl{2} &\qw &\qw       &\qw \\
        &\lstick{$r$} &\targ{}   &\qw       &\qw       &\ctrl{-1}            &\qw       &\qw       &\targ{}   &\qw\\
        &\lstick{$s$} &\qw       &\ctrl{-3}   &\targ{}      &\ctrl{-2}   &\targ{}     &\ctrl{-3}  &\qw       &\qw
    \end{quantikz}
    }\end{center}

    \begin{center}
    \noindent\makebox[\textwidth]{
    \begin{quantikz}
        &\lstick{$p$} &\ctrl{2} &\targ{}       &\qw   &\qw                 &\qw   &\targ{}       &\ctrl{2} &\qw\\
        &\lstick{$q$} &\qw       &\ctrl{-1}  &\ctrl{2} & \gate{R_y(\theta)} &\ctrl{2} &\ctrl{-1} &\qw       &\qw \\
        &\lstick{$r$} &\targ{}   &\qw       &\qw       &\ctrl{-1}            &\qw       &\qw       &\targ{}   &\qw\\
        &\lstick{$s$} &\qw       &\qw   &\targ{}      &\ctrl{-2}            &\targ{}       &\qw  &\qw       &\qw
    \end{quantikz}
    }\end{center}

    \section*{Appendix C. Get analytic energy function}

    To calculate the single parameter energy reduction for one operator $\tau$, one need to deal with the following minimization
    \begin{equation}
        \min_{\theta} E(\theta) = \min_{\theta} \bra{\psi_k} U^\dag(\theta) H U(\theta) \ket{\psi_k}
    \end{equation}
    where $U(\theta) = e^{\theta\tau}$, $\ket{\psi_k}$ represent the final state after the k-th iteration. It is expensive to calculate it using single parameter VQE. Here we provide an analytic and efficient method to deal with it, get the energy function using quantum computer and perform the minimization in classical computer.

    The exponential of an excitation operator in FEB, QEB and sQEB pools satisfy the expansion as below
    \begin{equation}
        U(\theta)=e^{\theta \tau}=1+\sin{\theta}\tau +(\cos{\theta}-1)(-\tau^2)
    \end{equation}
    This expansion allows us to calculate the energy $E(\theta)$. The energy function is
    \begin{align*}
        E(\theta)
        &= \bra{\psi_k} U^\dag(\theta) H U(\theta) \ket{\psi_k} \\
        &= \bra{\psi_k} (1 - \sin{\theta}\tau + (\cos{\theta}-1)(-\tau^2)) 
        H
        (1 + \sin{\theta}\tau + (\cos{\theta}-1)(-\tau^2)) \ket{\psi_k} \\
        &= \bra{\psi_k} H \ket{\psi_k} - \sin{\theta} \bra{\psi_k} (\tau H-H\tau) \ket{\psi_k} - (\cos{\theta}-1) \bra{\psi_k} (\tau^2H + H\tau^2) \ket{\psi_k} \\
        & -\sin^2{\theta} \bra{\psi_k} \tau H\tau \ket{\psi_k} + \sin{\theta} (\cos{\theta}-1) \bra{\psi_k} (\tau H\tau^2 - \tau^2H\tau) \ket{\psi_k} + (\cos{\theta}-1)^2 \bra{\psi_k} \tau^2H\tau^2 \ket{\psi_k}
        \\
        &= f_0 + f_1\sin{\theta} + f_2\sin{2\theta} + f_3\cos{\theta} + f_4\cos{2\theta}
    \end{align*}
    in which coefficients $f_0-f_4$ are expectation value of some observable in state $\ket{\psi_k}$.
    \begin{align}
        f_0 &= \bra{\psi_k} H + (\tau^2H + H\tau^2) -\frac12 \tau H\tau +\frac32 \tau^2H\tau^2 \ket{\psi_k} \\
        f_1 &= \bra{\psi_k} (H\tau-\tau H) - (\tau H\tau^2 - \tau^2H\tau) \ket{\psi_k} \\
        f_2 &= \bra{\psi_k} \frac12 (\tau H\tau^2 - \tau^2H\tau) \ket{\psi_k} \\
        f_3 &= \bra{\psi_k} -(\tau^2H + H\tau^2) -2 \tau^2H\tau^2 \ket{\psi_k} \\
        f_4 &= \bra{\psi_k} \frac12 \tau H\tau + \frac12 \tau^2H\tau^2 \ket{\psi_k}
    \end{align}
    One can measure these expectations to determine the analytic energy function, then the minimum point of this function can be obtained in classical computer easily.

    Actually, for operator pool employ Pauli strings as generator, the single-parameter $\Delta E$ can be obtained without measurement overhead compared to gradient calculation. 
    If $\tau = iP$, the Pauli string $P$ satisfy $P^2=I$ so $\tau^2 = -1$, so the coefficient $f_0-f_4$ can be simplified as below
    \begin{align}
        f_0 &= \bra{\psi_k} \frac12 H -\frac12 \tau H\tau \ket{\psi_k} \\
        f_1 &= 0 \\
        f_2 &= \bra{\psi_k} \frac12 (H\tau - \tau H) \ket{\psi_k} \\
        f_3 &= 0 \\
        f_4 &= \bra{\psi_k} \frac12 \tau H\tau + \frac12 H \ket{\psi_k}
    \end{align}
    So one need to measure 1) $\bra{\psi_k} H \ket{\psi_k}$ the expectation value of $\ket{\psi_k}$ which is already known, 2) $\bra{\psi_k} [H,\tau] \ket{\psi_k}$ the gradient of $\tau$ which is also need to be measured when using gradient criterion, 3) $\bra{\psi_k} \tau H \tau \ket{\psi_k}$ expectation value of $\tau H \tau$ with respect to $\ket{\psi_k}$.

    The term need extra attention for $\Delta E$ evaluation is only the expectation value of $\tau H \tau$, but we will see the Pauli strings in this operator is exactly identical to the Pauli strings in the hamiltonian $H$. So the expectation value of $\tau H \tau$ is obtained as long as the expectation value of $H$ is obtained, which is already measured after k-th iteration.

    Let's prove that $\tau H \tau$ have exactly identical Pauli strings as in $H$. 
    \begin{equation}
        H = \sum_i h_i P_i
    \end{equation}
    $\tau$ is a Pauli string, we know two Pauli string either commute or anti-commute, so $\tau P = (-1)^{n(\tau,P)} P \tau$, $n$ be either 0 (commute) or 1 (anti-commute), so we get
    \begin{align}
        \tau H \tau &= \sum_i h_i \tau P_i \tau \\
        &= \sum_i (-1)^{n(\tau,P_i)} h_i P_i \tau \tau \\
        &= \sum_i (-1)^{n(\tau,P_i)} h_i P_i
    \end{align}
    So $\tau H \tau$ have exactly identical Pauli strings as in $h$, only differ by the sign for some Pauli strings.
    
    So, the measurement cost for $\Delta E$ and gradient is identical for operator pools of Pauli strings.
\end{document}